\documentclass[preprint2]{aastex}
\usepackage{graphicx}

\slugcomment{Accepted for publication in the Astrophysical Journal}

\shorttitle{XMM-Newton and Chandra Observations of G352.7-0.1}
\shortauthors{Pannuti et al.}

\begin{document}

\title{{\it XMM-Newton} and {\it Chandra} Observations of the Ejecta-Dominated
Mixed-Morphology Galactic Supernova Remnant G352.7-0.1}

\author{Thomas G. Pannuti\altaffilmark{1},  Oleg 
Kargaltsev\altaffilmark{2}, Jared P. Napier\altaffilmark{1} and Derek Brehm\altaffilmark{2}}

\altaffiltext{1}{Space Science Center, Department of Earth and Space Sciences,
235 Martindale Drive, Morehead State University, Morehead, KY 40351; 
t.pannuti@moreheadstate.edu, jpnapier@moreheadstate.edu}
\altaffiltext{2}{Department of Physics, 308 Samson Hall, George Washington University,
Washington, DC 20052; kargaltsev@gwu.edu, brehm.derek@gmail.com}

\begin{abstract}

We present a spatial and spectral X-ray analysis of the Galactic supernova remnant (SNR)
G352.7-0.1 using archival data from observations made with the {\it XMM-Newton} X-ray Observatory
and the {\it Chandra} X-ray Observatory. Prior X-ray observations of this SNR had revealed a 
thermal center-filled morphology which contrasts with a shell-like radio morphology, thus
establishing G352.7$-$0.1 as a member of the class of Galactic SNRs known as mixed-morphology 
SNRs (MMSNR). Our study confirms that the X-ray emission comes from the SNR interior and must 
be ejecta-dominated. Spectra obtained with {\it XMM-Newton} may be fit satisfactorily with a single 
thermal component (namely a non-equilibrium ionization component with enhanced abundances of 
silicon and sulfur). In 
contrast, spectra extracted by {\it Chandra} from certain regions of the SNR cannot always be
fit by a single thermal component. For those regions, a second thermal component
with solar abundances or two thermal components with different temperatures and thawed silicon and 
sulfur abundances (respectively) can generate a statistically-acceptable fit. We argue that the former 
scenario is more physically-plausible: based on parameters of our spectral fits, we calculate physical 
parameters including X-ray-emitting mass ($\sim$45 $M$$_{\odot}$, for solar abundances). 
We find no evidence for overionization in the X-ray emitting plasma associated
with the SNR: this phenomenon has been seen in other MMSNRs. We have conducted a search for 
a neutron star within the SNR using a hard (2-10 keV) {\it Chandra} image but could not identify
a firm candidate. We also present (for the first time) the detection of infrared emission from 
this SNR as detected at 24 $\mu$m by MIPS aboard {\it Spitzer}. Finally, we discuss the properties of
G352.7$-$0.1 in the context of other ejecta-dominated MMSNRs. 
\end{abstract}


\section{Introduction}
\label{IntroductionSection}

The term ``supernova remnant" (SNR) refers to the expanding shell of material produced by a 
supernova explosion: SNRs are typically composed of stellar ejecta mixed with swept-up 
interstellar material (the latter component tends to dominate over the former component as the 
SNR ages). SNRs are intimately involved with many processes associated with the interstellar 
medium (ISM), such as chemical evolution (through the enrichment of heavy element content) 
and cosmic-ray acceleration. SNRs have also been detected over a remarkably broad portion of 
the electromagnetic spectrum ranging from low-energy radio emission (synchrotron radiation)
to TeV $\gamma$-rays. The observed high-energy emission can come from the SNR shell
(forward shock) which accelerates particles to very high energies and the SNR interior (filled with
heated ejecta and/or swept-up material and accelerated electrons). In some cases both
components are clearly present (composite type SNRs) while in the others only the shell
or only the interior emission (center-filled SNRs) could be identified. To date, approximately 280 
SNRs are known to exist in our galaxy \citep{Green09a,Green09b}: these sources have been detected 
chiefly by their radio emission. Based on their radio morphologies, about 78\% belong to the shell type,
12\% are composite type and 4\% are the center-filled type. The remaining 6\% are difficult
to attribute to any of the above types \citep{Green09a,Green09b}.
\par
While theories about the evolution of SNRs generally predict that these sources should have 
shell-like morphologies in the X-ray and radio, a particular class of SNRs known as 
mixed-morphology SNRs (MMSNRs) have garnered considerable interest from observational, 
theoretical and computer modeling perspectives \citep{Rho98}. These sources feature the nominal 
shell-like radio morphology typical of SNRs coupled with a thermal center-filled X-ray morphology
which is centrally-brightened and roughly homogeneous \citep{Vink12}. MMSNRs also appear
to be associated with molecular clouds, as evidenced by hydroxyl (OH) masers at 1720 MHz
that are indicators of 
interactions between SNR shocks and molecular clouds and by infrared observations that reveal
emission associated with shocked molecules. The origin of the contrasting X-ray and radio 
morphologies that characterize MMSNRs remains a mystery. Several competing theories have been
proposed to account for the unusual morphologies exhibited by MMSNRs: one theory is the 
evaporating cloud theory \citep{White91} which states that the interior of an MMSNR is filled with
clouds that have survived passage by the initial SNR blast wave. These clouds have thus loaded the 
interior of the SNR with a reservoir of material and produce thermal X-ray emission as they evaporate.
A second theory proposed by \citet{Cox99} suggests that the overfed thermal emission may be 
explained by efficient thermal conduction processes within the SNR interior. Unfortunately, neither of
these theories can by themselves provide an adequate explanation for the observed X-ray 
morphology of these particular SNRs. For example, \citet{Shelton04} noted that both the evaporating
cloud model and the thermal conduction model fail to adequately predict the sharp X-ray surface
brightness gradient observed for several MMSNRs. A pivotal clue about this phenomenon
may rest with the fact that MMSNRs all preferentially appear to be interacting
with adjacent molecular clouds. Remarkably, several MMSNRs appear
to feature X-ray-emitting plasma that is dominated by ejecta rather than swept-up material as 
expected for sources with ages estimated to be $\sim$10$^4$ yr (the typical estimated age of
an MMSNR). To better understand the 
X-ray properties of 
MMSNRs -- and specifically MMSNRs with ejecta-dominated
X-ray emission -- we present an analysis of X-ray observations of the Galactic 
ejecta-dominated MMSNR G352.7$-$0.1.  
\par
G352.7$-$0.1 was first classified as a SNR based on observations made at 408 and 5000 MHz of the
source by \citet{Clark73} and \citet{Clark75} using the Molonglo Cross Radio Observatory and the 
Parkes 64 Meter Radio Telescope, respectively. Subsequent radio observations of G352.7$-$0.1
\citep{Caswell83,Dubner93,Whiteoak96,Giacani09} with such observatories as the Fleurs Synthesis
Telescope, the Very Large Array (VLA) and the Molonglo Observatory Synthesis Telescope at 
frequencies of 843 MHz and 1400 MHz have revealed a shell-like morphology of this source 
that features two loops or lobes: the second lobe is an incomplete shell seen toward the southwest 
of the complete first lobe. The entire angular extent of the SNR is approximately 8$\times$6 
arcminutes \citep{Green09a} and the integrated spectral index of the SNR is $\alpha$ $\sim$ $-$0.6 
\citep{Dubner93} 
(where we have adopted the convention for radio flux densities $S$ $\propto$ $\nu$$^{\alpha}$).   
The assumed distances to G352.7$-$0.1 have ranged from 7.5$\pm$0.5 kpc \citep{Giacani09} to 
8.5 kpc \citep{Kinugasa98} to 11 kpc \citep{Dubner93} to 16.4 kpc \citep{Caswell83}. 
Based on observations that traced the HI absorption profile in the direction of G352.7$-$0.1 and
considering the Galactic tangent point velocity, \citet{Giacani09} derived bounds of between 
$\sim$6.8 kpc and $\sim$8.4 kpc for the distance to the SNR. Those authors concluded that a value 
of 7.5$\pm$0.5 kpc is
a satisfactory distance estimate to G352.7$-$0.1: we adopt this distance to the SNR in the present 
paper.  In Table \ref{G352PropsTable} we present a summary of the general physical properties of 
G352.7$-$0.1.
\par
A bright unresolved radio source is seen on the eastern edge of the complete first lobe and
\citet{Dubner93} speculated that this source may be a ``compact inhomogeneity in the radio shell" or 
an unrelated background source seen against the radio shell by a chance alignment. As pointed out by 
\citet{Giacani09}, this radio source was resolved into two separate point-like sources by higher angular 
resolution observations: these two sources have been cataloged as WBH2005 352.775$-$0.153 and 
WBH2005 352.772$-$0.149 in the ``New Catalog of Compact 6cm Sources in the Galactic Plane" 
\citep{White05}. To further elucidate the properties of these radio sources, \citet{Giacani09} computed 
a spectral index of the combined emission from the two sources (because these two sources were 
unresolved in the 1.4 GHz map those authors presented of G352.7$-$0.1), of $\alpha$ $\sim$ $-$1.1 
to $-$1.3. This estimate of $\alpha$ prompted \citet{Giacani09} to conclude that these two sources 
comprise clouds of radio-emitting plasma ejecta by a distant active galactic nuclei seen in projection 
beyond G352.7$-$0.1 and thus physically unrelated to the SNR. Therefore, we do not consider these 
radio sources to be associated with G352.7$-$0.1. We note that \citet{Green97} and 
\citet{Koralesky98} detected OH maser emission at 1720 MHz from G352.7$-$0.1: 
such emission is closely associated with interactions between SNRs and molecular clouds.
MMSNRs (like G352.7$-$0.1) appear to be interacting with adjacent molecular clouds 
and this interaction is suspected to at help explain the origin of the observed contrasting X-ray 
and radio morphologies of these sources. Additional evidence for an interaction between 
G352.7$-$0.1 and adjacent molecular clouds is supplied by \citet{ToledoRoy14}, who explain 
the observed radio morphology using a model where the original supernova explosion occurs inside
and near a molecular cloud. 
\par
X-ray emission from G352.7$-$0.1 was first detected by \citet{Kinugasa98} based on observations 
made with the Advanced Satellite for Cosmology and Astrophysics ({\it ASCA}) \citep{Tanaka94} as 
part of the {\it ASCA} Galactic Plane Survey (see \citet{Sugizaki01}, who also reported the detection of  
X-ray emission from G352.7$-$0.1). Spectral analysis of this SNR as performed by \citet{Kinugasa98} 
using extracted spectra from the Gas Imaging Spectrometers (GISs) and the Solid State Imaging 
Spectrometers (SISs) aboard {\it ASCA} 
(see \citet{Burke94}, \citet{Serlemitsos95}, \citet{Makishima96} and \citet{Ohashi96} for more 
information about the GISs, the SISs and the XRTs) revealed that the X-ray emitting plasma could be 
best fit with a non-equilibrium ionization (NEI) model with an ionization timescale $\tau$ $\sim$ 
10$^{11}$ cm$^{-3}$ s (indicating that the plasma is not in ionization equilibrium) and a temperature of 
$kT$ $\sim$ 2 keV which is unusually high for an SNR. Prominent K-shell lines from highly-ionized 
silicon, sulfur and argon were also clearly seen in the extracted spectra: the fitted abundances of 
silicon and sulfur were both overabundant relative to solar abundances ($\sim$3.7 and $\sim$3.4, 
respectively) and thus the X-ray-emitting plasma associated with G352.7$-$0.1 is ejecta-dominated.
Lastly, \citet{Kinugasa98} -- who assumed that G352.7$-$0.1 lies at a distance of 8.5 kpc and that the
SNR is in the Sedov stage of evolution -- estimated the age of the SNR to be $t$ $\sim$ 2200 yr. If 
this age estimate is accurate, then it is remarkable that the X-ray emitting plasma associated with 
G352.7$-$0.1 is still ejecta-dominated. We note that a separate spectral analysis of the {\it ASCA}-
detected emission from G352.7$-$0.1 was conducted by \citet{Reynolds99}, who fit the extracted 
spectra using the synchrotron X-ray emission model SRCUT: this model assumes that all of the 
observed X-ray continuum emission is produced by synchrotron radiation and is characterized by the
radio flux density of the SNR at 1 GHz, the radio spectral index $\alpha$ and the roll-off frequency
$\nu$$_{\rm{rolloff}}$ from which the energy $E$$_{\rm{max}}$ (the energy at which the electron
energy distribution must steepen from its slope at radio-emitting energies). Based on the results of
their fits to the spectra (which were constrained by the observed radio properties of G352.7$-$0.1),
those authors estimated a value of 40 TeV for $E$$_{max}$. 
\par
An analysis of a pointed X-ray observation of G352.7$-$0.1 with the {\it XMM-Newton} 
Observatory \citep{Turner01} is presented by \citet{Giacani09}. The pn-CCD camera \citep{Struder01} 
and the Multi-Object Spectrometer cameras \citep{Turner01} (hereafter referred to as the PN, MOS1 
and MOS2 respectively for the remainder of this paper) aboard {\it XMM-Newton} comprise the 
European Photon Imaging Camera (EPIC). Based on the superior 
resolution attained by the PN, MOS1 and MOS2 in imaging G352.7$-$0.1, \citet{Giacani09} disagreed 
with \citet{Kinugasa98} (who had concluded that the X-ray morphology of G352.7$-$0.1 was shell-like, 
similar to its radio morphology) and instead classified the X-ray morphology as center-filled, therefore
motivating a classification of G352.7$-$0.1 as an MMSNR. \citet{Giacani09} conducted a spectral 
analysis of extracted PN, MOS1 and MOS2 spectra for the entire angular extent of G352.7$-$0.1: the 
extracted spectra were fit with an NEI model (the same model implemented by \citet{Kinugasa98}) and 
the parameters of the fit to the spectra were broadly similar to those reported by \citet{Kinugasa98}. 
Specifically, \citet{Giacani09} reported fit values of $kT$ $\sim$ 1.9$\pm$0.2 keV, $\tau$ = 4.5$\pm
$0.5$\times$10$^{10}$ cm$^{-3}$ s and abundances for silicon and sulfur of 2.4$\pm$0.2 and 
3.8$\pm$0.3 relative to solar, respectively. \citet{Giacani09} also reported an overabundance of argon 
(4.7$\pm$1.2 relative to solar) in the X-ray emitting plasma associated with G352.7$-$0.1, the first 
report of an overabundance of argon in the X-ray emitting plasma associated with this SNR. 
\citet{Giacani09} also commented on the presence of an Fe K emission line at 6.46$\pm$0.03 keV in 
the extracted spectra. Based on the fit values of these parameters and assuming a distance to this
SNR of 7.5 kpc, \citet{Giacani09} estimated the electron density of the plasma to be $n$$_e$ = 0.3 
cm$^{-3}$, the age of the SNR to be $t$ $\sim$ 4700 yr,  
an X-ray emitting mass of 10 M$_{\odot}$ and an explosion energy of 10$^{50}$ ergs.   
Lastly, \citet{Giacani09} found no X-ray counterparts to the radio sources WBH2005 352.775$-$0.153 
and WBH2005 352.772$-$0.149. 
\par
In the present paper we present an imaging and spectroscopic analysis of the properties of 
G352.7$-$0.1 based on pointed X-ray observations made by {\it XMM-Newton} and the 
{\it Chandra} X-ray Observatory: the present work contains (to the best of our knowledge)
the first analysis of the dataset from the {\it Chandra} observation of this SNR. The {\it XMM-Newton} 
and {\it Chandra} observations of G352.7$-$0.1, as well as the accompanying data reduction, are 
presented and described in Sections \ref{XMMObsDataRed} and \ref{ChandraObsDataRed}, 
respectively. In Sections \ref{XMMG352} and \ref{ChandraG352} we present spatial and spectral 
analysis of the {\it XMM-Newton} and {\it Chandra} datasets, respectively. We discuss properties
of this SNR in Section \ref{DiscussionSection}, where we compare our results with 
previously-published analysis of the SNR (in Section \ref{InterpretationSubSection}). We also 
calculate and discuss 
physical properties of G352.7$-$0.1 -- including electron density, pressure and X-ray-emitting mass --
in Section \ref{PhysicalPropertiesSubSection}. In Section \ref{OverionizationSubSection} we 
investigate whether the X-ray-emitting plasma associated with G352.7$-$0.1 is overionized (as seen 
in the cases of other Galactic  MMSNRs). A search for a neutron star associated with 
this SNR through an analysis of the spectral properties of hard unresolved X-ray sources seen in
projection toward the interior of the SNR is described in Section
\ref{UnresolvedXraySourceSubSection}. In Section \ref{IRG352SubSection} we present the discovery 
of infrared emission from G352.7$-$0.1 as observed by the {\it Spitzer Space Telescope} and in
Section \ref{EjectaDominatedMMSNRSubSection} we discuss this SNR within the context of other
MMSNRs that feature ejecta-dominated X-ray-emitting plasmas.
Finally, our conclusions are presented in Section \ref{ConclusionsSection}.

\section{Observations and Data Reduction}

\subsection{{\it XMM-Newton} Observations and Data Reduction\label{XMMObsDataRed}}

G352.7$-$0.1 was the subject of a pointed {\it XMM-Newton} observation made 
on 2002 October 3 with all three EPIC cameras: the Sequence Number of the observation was 
0150220101 (PI: J. Hughes). The fields of view of the MOS1, MOS2 and 
PN are all approximately 30 arcminutes: therefore the entire angular extent of
G352.7$-$0.1 was easily sampled by a single pointed observation.  The observation was
conducted with all three cameras in MEDIUM filter mode: in addition, the PN camera was in Full
Frame mode. The angular resolution of the MOS1, MOS2 and PN
are all approximately 6 arcseconds at 1 keV: at the same energy, the effective collecting areas are 922 
cm$^2$ for the two MOS cameras and 1227 cm$^2$ for the PN camera. These three cameras are
each nominally sensitive to photons with energies between 0.2 and 12.0 keV. 
\par
Data for the observation of G352.7$-$0.1 was obtained from the on-line {\it XMM-Newton} data
archive\footnote{See http://xmm.esac.esa.int/xsa/.}. To reduce the data, standard tools in the 
HEAsoft software package\footnote{See http://heasarc.gsfc.nasa.gov/.} (Version 6.12) and the 
Science Analysis System (SAS -- \citet{Gabriel04}) software 
package\footnote{See http://xmm.esa.int/sas/.} (Version 12.0.0) were used. Specifically, the 
SAS tools $\tt{emchain}$ and $\tt{mos-filter}$ were used to process data from the observations 
made by the MOS1 and the MOS2 cameras and to identify good-time intervals within the
observations. Likewise, the SAS tools $\tt{enchain}$ and $\tt{pn-filter}$ were used to process data 
from the observation made by the PN camera and to identify good time intervals within the 
observation. After filtering, the effective exposure times for the MOS1, MOS2 and PN observations of 
G352.7$-$0.1 were 28639 s, 28650 s and 16468 s, respectively. We extracted MOS1, MOS2 and PN 
spectra for both the entire SNR as well as the bright eastern region of the SNR. To perform the
spectral extraction for both the source and background regions, we used the SAS tool 
$\tt{evselect}$ and the extracted spectra were grouped to a minimum of 25 counts per channel.
Subsequently, the SAS tool $\tt{backscale}$ was used to compute the proper value for the BACKSCAL 
keyword in the extracted spectra. Finally, the SAS tools $\tt{rmfgen}$ and $\tt{arfgen}$ were used to
generate the response matrix files (RMFs) and the ancillary response files (ARFs), respectively, for the 
spectra required for spectral fitting. An analysis of the {\it XMM-Newton} extracted
spectra is presented in Section \ref{XMMG352}.
In Table \ref{XMMObsTable} we present a summary of the
{\it XMM-Newton} observation of G352.7$-$0.1.

\subsection{{\it Chandra} Observations and Data Reduction\label{ChandraObsDataRed}}

G352.7$-$0.1 was the subject of a pointed observation made with the {\it Chandra} X-ray Observatory
\citep{Weisskopf02} on 2004 October 6. The corresponding ObsID of this observation is 4652 
(PI: J. Hughes) and the observation was made in VERY FAINT timed mode with the Advanced CCD 
Imaging Spectrometer (ACIS) at a focal plane temperature of $-$118.54 C. The ACIS is comprised of 
the ACIS-S and ACIS-I arrays of chips: each chip is approximately 8.3$\arcmin$ $\times$ 8.3$\arcmin$ 
(see \citet{Garmire03} for more details about the properties of the ACIS chips). The entire angular 
extent of the X-ray emitting plasma associated with G352.7$-$0.1 fits within the field of view of one of 
the ACIS chips (specifically the ACIS-S3 chip). This particular chip (which is back-illuminated) is
sensitive to photons with energies ranging from 0.3 to 10.0 keV and the FWHM angular resolution of 
this camera at 1 keV is 0.5$^{\arcsec}$. Finally, the effective
collecting area of the ACIS-S3 chip for photons with energies of 1.0 keV is 525 cm$^{2}$.  
\par
Data for the {\it Chandra} observation was obtained from the on-line {\it Chandra} data 
archive\footnote{See http://cda.harvard.edu/chaser/.}: to reduce the data, standard tools in the 
{\it Chandra} Interactive Analysis of Observations (CIAO\footnote{See \citet{Fruscione06} and
http://cxc.harvard.edu/ciao/.}) software package Version 4.4 (CALDB Version 4.4.7) was used 
to reduce and analyze the data. The CIAO tool $\tt{chandra\_repro}$ was run to reprocess
the dataset and to perform all of the recommended data processing steps: these steps included the 
application of the latest temperature-dependent charge transfer inefficiency correction, the latest 
time-dependent gain-adjustment and the latest gain map. The tool also flags bad pixels (creating a
new bad pixel file) and applies additional filters based on grades and proper status columns. Running
this tool creates a new EVT2 file which we further filtered to exclude time periods of significant background
flare activity. This particular filtering was accomplished using the CIAO tool 
$\tt{chips}$\footnote{See http://cxc.harvard.edu/chips/index.html.}: after all of the filtering was applied,
the effective exposure time of the processed EVT2 file was 44583 seconds. In Table 
\ref{ChandraObsTable} we present a summary of the {\it Chandra} observation of G352.7$-$0.1.  
\par
To perform a spectral analysis of the X-ray emitting plasma associated with G352.7-0.1, the tool 
$\tt{specextract}$ was used to generate spectral files -- source spectra, background spectra, ARFs and
RMFs -- of both the entire SNR as well as several regions within the SNR. The extracted spectra were
grouped to a minimum of 15 counts per channel. An analysis of all of the
{\it Chandra}-extracted spectra is presented in Section \ref{ChandraG352}. To reduce the effect of
point source contamination of the extracted spectra, the CIAO tool $\tt{wavdetect}$ -- a wavelet-based 
source algorithm \citep{Freeman02} -- was used to identify unresolved sources detected by the 
{\it Chandra} observation and exclude them while analyzing the diffuse emission (as described
below). 

\section{Analysis and Results}

\subsection{{\it XMM-Newton} Observations of G352.7$-$0.1: Imaging and 
Spectroscopy\label{XMMG352}}

\begin{figure}
\plotone{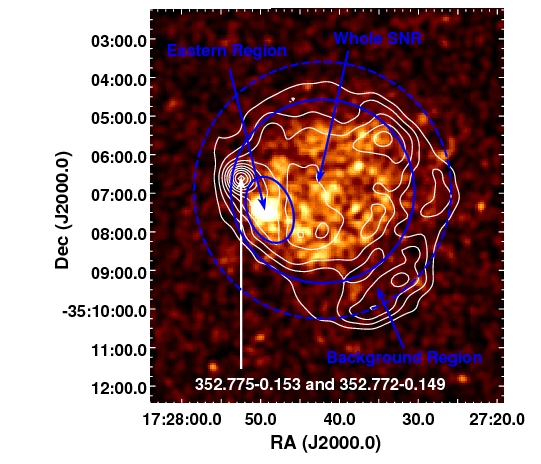}
\caption{\small{{\it XMM-Newtwon} MOS1 image (in color) of G352.7$-$0.1: the image depicts 
emission from the SNR as detected from the energy range between 0.5 through 10.0 keV. The 
brightness range of the image is 0 to 0.15 counts s$^{-1}$ arcsec$^{-2}$. Gaussian smoothing with 
a radius of 10 arcseconds has been applied. The white contours depict radio emission as detected 
by the Very Large Array (VLA) at a frequency of 1.4 GHz \citep{Dubner93,Giacani09}: the contour 
levels range from 
0.02 to 0.11 Jy/beam in steps of 0.01 Jy/beam. The location of the background double radio sources
WBH 2005 352.775$-$0.153 and WBH 2005 352.772$-$0.149 are indicated. The regions of spectral
extraction -- namely for the bright eastern region, the whole SNR and the background region -- are
also indicated. Notice that the X-ray morphology is center-filled but the radio morphology is 
shell-like. These contrasting morphologies are the defining characteristics of MMSNRs. See Section 
\ref{XMMG352}. \label{G352_XMM_Radio}}}
\end{figure}

In Figure \ref{G352_XMM_Radio} we present an {\it XMM-Newton} MOS1 image of G352.7$-$0.1. The 
image depicts emission detected for the energy range of 0.5 to 10.0 keV. The emission has been 
smoothed with a Gaussian with a radius of 10 arcseconds. Radio emission as detected by the Very 
Large Array (VLA) at a frequency of 1.4 GHz \citep{Dubner93,Giacani09} is depicted with overlaying 
contours (this radio map was provided courtesy of Gloria Dubner and was published in 
\citet{Dubner93}). The combined position of the 
background double radio sources WBH 2005 352.775$-$0.153 and WBH 2005 352.772$-$0.149 
(as seen toward the eastern boundary of the main radio shell) is indicated in the figure. No significant 
X-ray emission is detected from these background radio sources \citep[in agreement with
the results of][]{Giacani09}. In addition, no significant X-ray emission is detected from the radio lobe of
the SNR seen toward the southwest. As noted by \citet{Giacani09}, the X-ray emission from 
G352.7$-$0.1 is seen inside the main radio shell, thus motivating its classification as a 
MMSNR. The X-ray emitting plasma appears to be broadly uniform in brightness with
the exception of a bright region seen toward the eastern edge. 
\par
\begin{figure}
\includegraphics[angle=-90,scale=0.7]{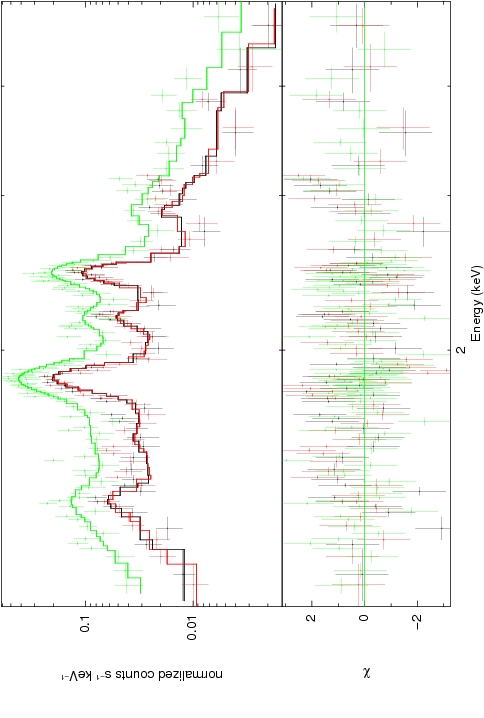}
\caption{\small{Extracted {\it XMM-Newton} MOS1, MOS2 and PN spectra (in black, red and green,
respectively) for the entire diffuse emission of G352.7$-$0.1 (``whole SNR") as fitted with a 
PHABS$\times$VNEI
model with variable silicon and sulfur abundances. Notice the prominent silicon and sulfur lines at
approximately 1.7 keV and 2.3 keV, respectively. The parameters of the fit to the spectrum are
given in Table \ref{XMMSpectralTable}. See Section \ref{XMMG352}. \label{TotalXMMSpectra}}}
%\citep[see][]{heiles03}.
\end{figure}

MOS1, MOS2 and PN spectra were extracted from both the entire angular extent of the plasma as 
well as for the bright eastern region. The extraction region for the whole SNR was circular and 
approximately 2.4$\arcmin$ in radius while the extraction region for the bright eastern region of the 
SNR was elliptical with semi-major and semi-minor axes of 1.3$\arcmin$ and 0.6$\arcmin$, 
respectively. The background spectrum was extracted from an annular region concentric with the 
extraction region for the whole SNR. This annular region had an inner radius corresponding to the 
radius of the source region for the whole SNR and an outer radius of 3.3$\arcmin$. The locations of 
all of these spectral extraction regions are depicted in Figure \ref{G352_XMM_Radio}. The three 
spectra for the whole SNR were fitted simultaneously using the thermal model VNEI: this model is a 
non-equilibrium collisional plasma
model which assumes a constant temperature, a single ionization parameter and variable elemental
abundances \citep{Hamilton83, Liedahl95, Borkowski01}. The version of the VNEI model used in this
paper (Version 1.1) implements updated calculations of ionization fractions using dielectronic 
recombination
rates as provided by \citet{Mazzotta98}. The choice of the VNEI model (instead of the NEI model with 
elemental abundances frozen to solar values) is motivated by the prominent lines of silicon and sulfur 
seen 
in the spectra (as shown in Figure \ref{TotalXMMSpectra} as well as previous X-ray spectral analysis 
conducted of this SNR by \citet{Kinugasa98} and \citet{Giacani09}). The VNEI model was multiplied by
the PHABS model to account for photoelectric absorption along the line of sight to G352.7$-$0.1. 
\par
\begin{figure}
\includegraphics[angle=-90,scale=0.25]{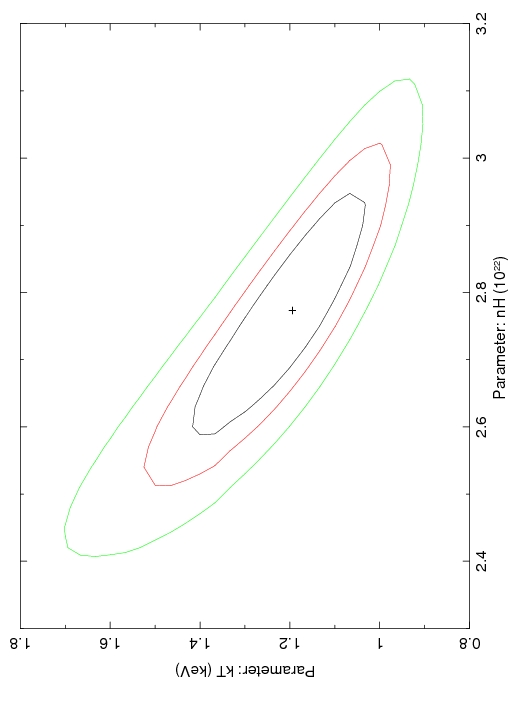}
\includegraphics[angle=-90,scale=0.25]{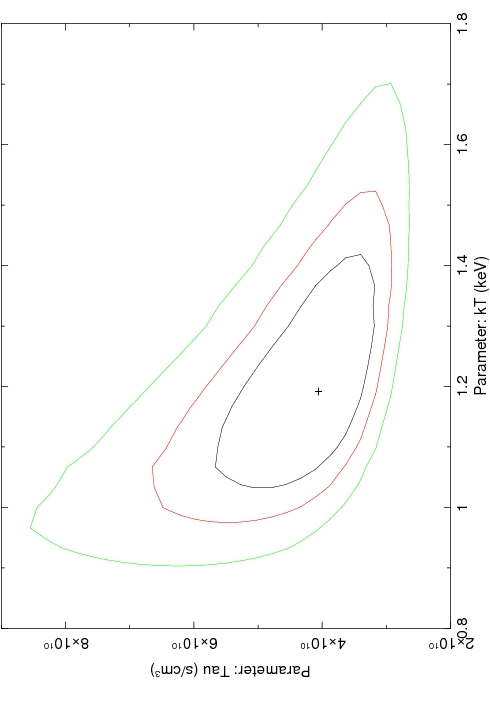}
\includegraphics[angle=-90,scale=0.25]{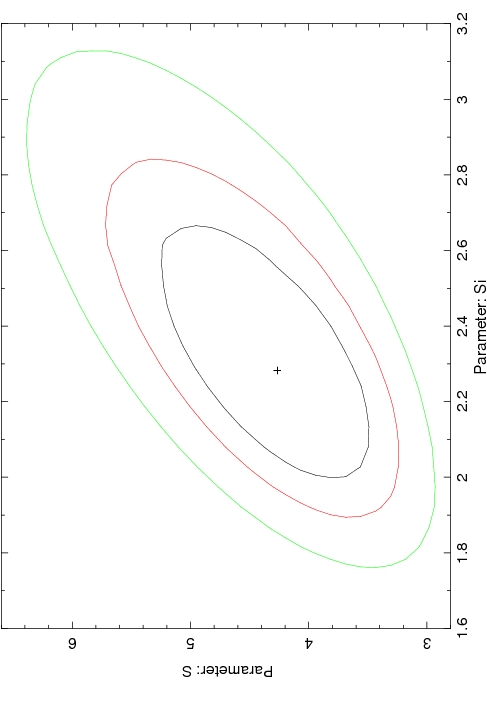}
\caption{\small{Confidence contour plots ($N$$_H$ vs. $kT$, $kT$ vs. $\tau$ and Si vs. S) for the 
PHABS$\times$VNEI fit to the extracted {\it XMM-Newton} MOS1+MOS2+PN spectra. 
The confidence contours are at the 1$\sigma$, 2$\sigma$ and
3$\sigma$ levels. See Table \ref{XMMSpectralTable} for the fit parameters and Section 
\ref{XMMG352} for a discussion of the fit.\label{TotalXMMSpectraConfPlots}}}
\end{figure}

We obtain statistically-acceptable fits (with $\chi$$_{\nu}^2$ values of $\sim$1.2 or less) to the 
extracted MOS1+MOS2+PN spectra for both the whole SNR as well as the bright eastern region.
In Table \ref{XMMSpectralTable} we present a summary of the fit parameters for the fits to both 
regions:
we note that the parameters of our fit for the spectra of the whole SNR are broadly similar to those
presented by \citet{Giacani09}. Regarding the
possible presence of spatial variations in the spectral properties, we find little 
difference between the derived fit parameters for the whole SNR compared to the 
bright eastern region. Therefore, this region appears to be simply a bright region of the thermal 
X-ray-emitting 
plasma instead of a pulsar wind nebula of a neutron star associated with G352.7$-$0.1 as one
could assume based on the images. The lack of obvious spatial variations in the spectral properties of 
the plasma is 
also consistent with the results of \citet{Giacani09}, who searched and failed to detect any such 
variations. In Figure \ref{TotalXMMSpectra} we present the extracted MOS1+MOS2+PN
spectra for the whole SNR (using the extraction regions for the source spectrum and the background 
spectrum as shown in Figure \ref{G352_XMM_Radio}). In Figure \ref{TotalXMMSpectraConfPlots} we
present confidence contour plots for the fit with the PHABS$\times$VNEI fit to the spectra of the whole
SNR. 

\subsection{{\it Chandra} Observations of G352.7$-$0.1: Imaging and Spectroscopy
\label{ChandraG352}}

\begin{figure}
\plotone{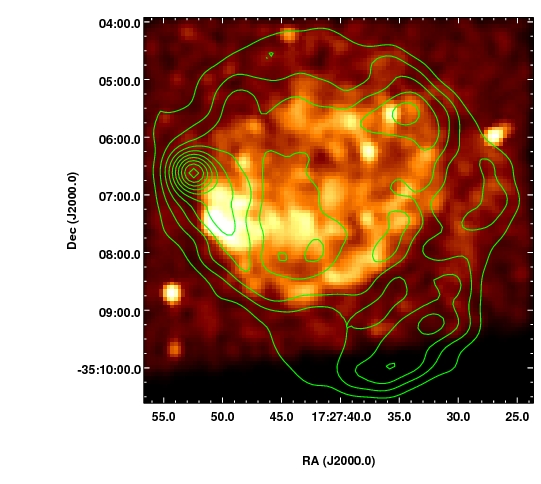}
\caption{Broadband (0.5-7.0 keV) {\it Chandra} image of G352.7$-$0.1 as observed with the 
ACIS-S3 chip: the chip gap can be clearly seen toward the bottom of the figure. The emission has
been smoothed with a Gaussian with a radius of 1.5 arcseconds. 
The green contours depict radio emission as detected by the VLA at a frequency of 1.4 GHz
and are placed at the same levels as in Figure \ref{G352_XMM_Radio}. See Section 
\ref{ChandraG352}.\label{G352_Chandra_Radio}} 
\end{figure}

\begin{figure}
\epsscale{1.00}

\includegraphics[scale=0.33]{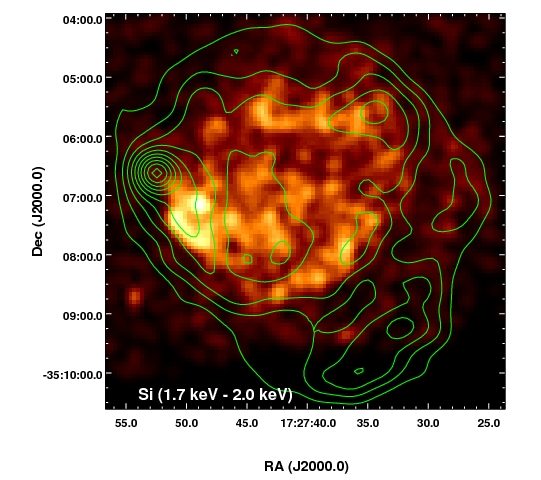}
\includegraphics[scale=0.33]{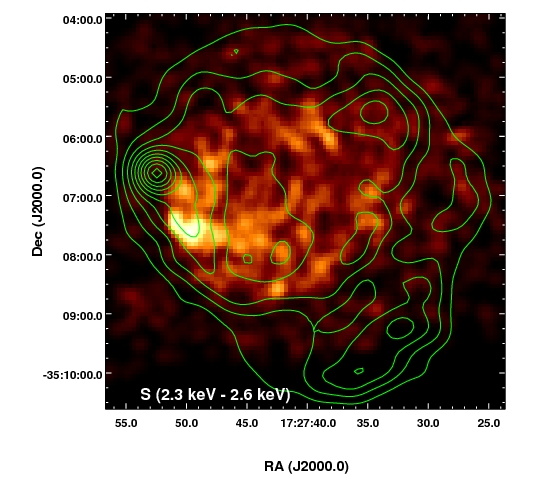}
\includegraphics[scale=0.33]{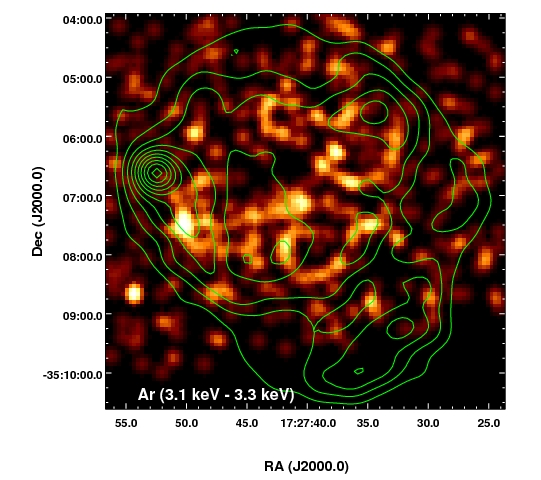}
\caption{\small{Narrowband {\it Chandra} images of G352.7$-$0.1 depicting X-ray emission as
produced by different elemental species, specifically (from left to right), silicon (1.7 keV - 2.0 keV),
sulfur (2.3 keV - 2.6 keV) and argon (3.1 keV - 3.3 keV). In each case, the emission has been
smoothed with a Gaussian with a radius of 1.5 arcseconds and green contours depicting radio 
emission as detected by the VLA at a frequency of 1.4 GHz are overlaid. The radio contours have
been placed at the same levels as seen in Figure \ref{G352_XMM_Radio}.  
\label{ChandraElementFigures}}}
\end{figure}

\begin{figure}
\plotone{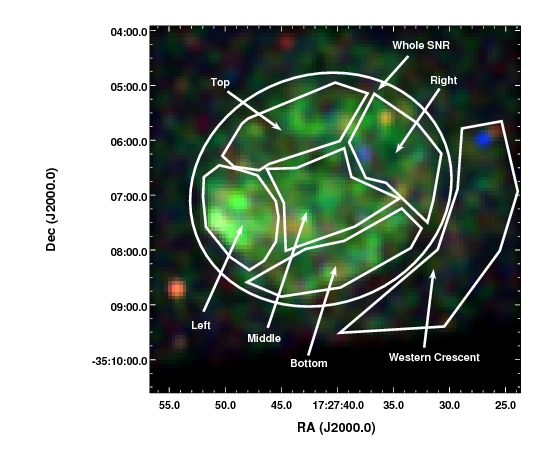}
\caption{Three-color {\it Chandra} image of G352.7$-$0.1: the depicted red, green and blue color 
scales correspond to soft (1.0-1.6 keV), medium (1.6-2.4 keV) and hard (2.4-7.0 keV) emission,
respectively. The emission at each energy range has been smoothed with a Gaussian with a
radius of 1.5 arcseconds. Regions of extraction of the diffuse emission for spectral analysis are 
indicated: the corresponding background spectrum is located just to the north of the source and out 
of the field of view. The entire source and the background region are contained within the field of view 
of the ACIS-S3 chip. See Section \ref{ChandraG352} and 
Tables \ref{ChandraPHABSVNEISpectralTable}, \ref{ChandraPHABSVNEINEISpectralTable} and
\ref{ChandraPHABSVNEIVNEISpectralTable}. \label{ChandraSpectralExtractionRegionsFig}}
\end{figure}

The superior angular resolution of the {\it Chandra} ACIS-S compared to the {\it XMM-Newton} 
MOS1, MOS2 and PN instruments allows us to perform a more detailed 
spatially-resolved X-ray spectroscopic study of G352.7$-$0.1. Our {\it Chandra} image of this
SNR is presented in Figure \ref{G352_Chandra_Radio} and
in Figure \ref{ChandraElementFigures} we follow the example of \citet{Giacani09} and present images 
made of G352.7$-$0.1 at three different energy cuts -- namely 1.7-2.0 keV, 2.3-2.6 keV and 3.1-3.3 
keV -- that correspond to known emission lines of silicon, sulfur and argon, respectively. We first 
extracted a spectrum for the entire SNR: we then divided the angular extent of the SNR 
into six different regions and extracted spectra from each region. Based on their apparent 
locations within the angular extent of G352.7$-$0.1, we denoted these smaller regions as 
``bottom," ``right," ``middle," ``left,"  ``top"  and the ``crescent" region (this latter region 
corresponds to emission seen in the space between the two radio loops seen to cleave at
the western portion of the SNR). We note that the ``left" region contains most of the ``eastern region"
considered previously in our {\it XMM-Newton} spectral analysis. A background spectrum was 
obtained using an elliptical region 
located north of the angular extent of G352.7$-$0.1 but still on the ACIS-S3 chip. The locations of the 
regions of spectral extraction are all shown in Figure 
\ref{ChandraSpectralExtractionRegionsFig}. All of the spectra were extracted using the CIAO tool 
{\tt specextract}: this tool generates all required spectral files, including the source spectra, 
background spectra and weighted ARFs and RMFs. When source and background spectral
files were extracted, care was taken to exclude counts from the locations of unresolved sources
(primarily background galaxies) found within the extraction regions. This was accomplished by 
running the CIAO source detection tool {\tt wavdetect}.
The locations of the sources detected by
{\tt wavdetect} were noted and counts from these locations were excluded when spectra
were extracted. We will return to the point sources detected by {\tt wavdetect} in Section
\ref{UnresolvedXraySourceSubSection} when we discuss the search for an X-ray-emitting neutron 
star associated with this SNR. 
\par
The spectra of all of the regions were fit with the same VNEI model (combined with a
PHABS model for the photoelectric absorption) employed to fit the extracted {\it XMM-Newton} 
spectra as described in the previous section. Again, the abundances of silicon and sulfur in the 
VNEI model were thawed while the abundances of the other elements were left frozen to solar 
values. 
In Tables \ref{ChandraPHABSVNEISpectralTable}, \ref{ChandraPHABSVNEINEISpectralTable}
and \ref{ChandraPHABSVNEIVNEISpectralTable} we summarize the results of our fits to all of the 
extracted {\it Chandra} spectra. In contrast to our results for the {\it XMM-Newton} 
spectra, a single VNEI component with enhanced silicon and sulfur abundances {\it cannot}
adequately fit all of the extracted {\it Chandra} spectra: unsatisfactory fits (with values for 
$\chi$$_{\nu}^2$ of more than 1.2) were obtained for the spectra of two of the smaller 
regions (namely the ``top" and ``left" regions) as well as the spectrum for the whole SNR. A 
single VNEI component with the enhanced silicon and sulfur abundances is sufficient to fit the
spectra of the ``bottom," ``right," ``middle" and ``crescent" regions. Inspection of the values
for the parameters of the fits reveals a significant variation in spectral properties across the whole
angular extent of the SNR. For example, we note that the fitted temperatures range from $kT$ 
$\sim$ 0.8 keV for the ``bottom" region to $kT$ $\sim$ 2.1 keV for the ``crescent" region. We
also note that the fitted abundances of silicon and sulfur show a range of values as well: in 
every region we considered the abundances for these two elements are super-solar. In contrast
to the work of \citet{Giacani09}, we find no evidence for an overabundance of argon in the spectra
of any of the regions that we considered: when we allowed the abundance of argon to vary, there
were no improvements in any of the values of $\chi$$_{\nu}^2$ for any of the fits to any of the regions.
We speculate that the {\it Chandra} observation of this SNR (unlike the {\it XMM-Newton} observation)
did not detect enough photons with energies corresponding to the argon line feature to investigate
the putative overabundance of this element in detail.
\par
\begin{figure}
\includegraphics[angle=-90,scale=0.7]
{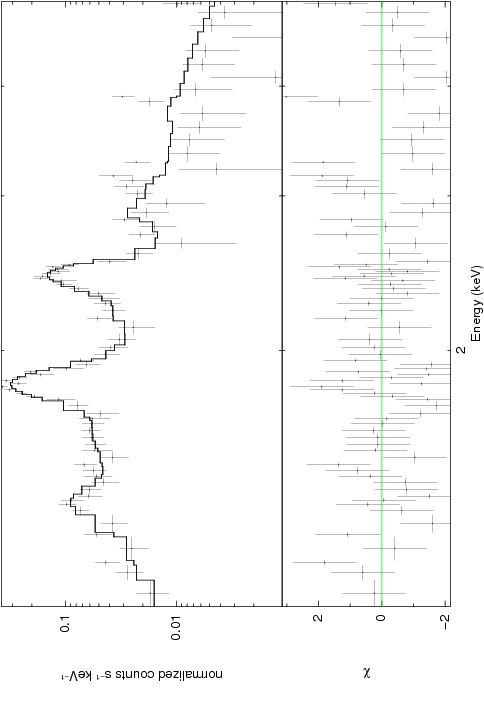}
\caption{Extracted {\it Chandra} ACIS spectrum of the entire diffuse emission of G352.7$-$0.1
as fitted with a PHABS$\times$(VNEI+NEI) model. See Section \ref{ChandraG352} and Table 
\ref{ChandraPHABSVNEINEISpectralTable}.\label{ChandraSpectrum}}
\end{figure}

To obtain statistically acceptable fits to the spectra of the ``left" region, the ``top" region
and the whole SNR, we added additional components to the spectral fits. We first added a single
power law component but we failed to obtain statistically acceptable fits (with $\chi$$^2_{\nu}$
$\leq$ 1.3) to the spectra of the ``left" region, the ``top" region nor the whole SNR. We then 
added an NEI component to our PHABS$\times$VNEI model: we left the abundance of this NEI 
component frozen to unity.
As an alternative, we also tried fitting the spectrum with a PHABS$\times$(VNEI+VNEI) model. 
In this case, we thawed the
abundance of silicon in one VNEI component and thawed the abundance of sulfur in the
other VNEI component (the abundances of all other elements were left frozen at values of
unity). In both cases, with the addition of a second thermal component, we were able to
obtain statistically acceptable fits to the spectra of the two individual regions as well as the
SNR as a whole. In Tables \ref{ChandraPHABSVNEINEISpectralTable} and
\ref{ChandraPHABSVNEIVNEISpectralTable} we present the results of these fits to the
extracted spectra for these regions. In Figure \ref{ChandraSpectrum} we present the {\it Chandra} 
spectrum of the diffuse emission of G352.7$-$0.1 as fit by the PHABS$\times$(VNEI+NEI) model.
We believe that the presence of this second component has
only manifested itself through the {\it Chandra} observation of G352.7$-$0.1 (but not the 
{\it XMM-Newton} observation of the SNR) due to the significantly lower background inherent
in observations made with the ACIS. 

\section{Discussion}
\label{DiscussionSection}

\subsection{Physical Interpretation}
\label{InterpretationSubSection}

We comment on the physical plausibility of these two models for interpreting the
spectra of these regions. Regarding the fits obtained with the PHABS$\times$(VNEI+VNEI) 
models, we note that in each case the sulfur-rich thermal component is implied to be
close to thermal equilibrium while the silicon-rich thermal component is not. We are not
aware of a physical mechanism where two different X-ray-emitting plasmas which are
enriched in different elemental abundances featuring different ionization timescales 
co-existing with each other. We therefore regard this model as not physically plausible.
In contrast, the PHABS$\times$(VNEI+NEI) may be readily explained by a plausible model
where the VNEI model fits emission from stellar ejecta while the NEI model fits emission from
swept-up material. Certainly, the large normalization of this latter component implies that the 
amount of swept-up material associated with this SNR is enormous. We note that a large swept-up
mass may be expected for MMSNRs like G352.7$-$0.1 which preferentially appear 
to be located near molecular clouds and appear to be interacting with these clouds as well.
We quantify this amount of mass
when we discuss the physical properties of G352.7$-$0.1 in the next section.

\subsection{Physical Properties of G352.7$-$0.1}
\label{PhysicalPropertiesSubSection}

We now discuss the physical properties of G352.7$-$0.1 as implied by the parameters derived
in fits to the extracted {\it Chandra} spectrum for the whole SNR. In our analysis we will follow the
approaches applied by \citet{Pannuti10} and \citet{Pannuti14} to other X-ray-detected Galactic 
MMSNRs (namely HB 21 (G89.0$+$4.7), CTB 1 (G116.9$+$0.2), Kes 17 (G304.6$+$0.1),
G311.5$-$0.3, G346.6$-$0.2 and CTB 37A (G348.5$+$0.1)). We first note that the normalizations 
$\mathcal{N}$ of the thermal models are defined by the relation

\begin{equation}
\mathcal{N} = \frac{10^{-14}}{4 \pi d^2} \int n_e n_p~dV 
\label{ThermalNormEqn}
\end{equation}

where $d$ is the distance to G352.7$-$0.1 (assumed to be 7.5 kpc), $n$$_e$ and $n$$_p$ are the 
electron and proton number densities (respectively) and $V$ is the volume of the SNR. We will 
assume that $n$$_e$ = 1.2 $n$$_p$ and that the filling factor of the 
X-ray-emitting gas is unity (based on the uniform morphology of the diffuse emission). From this 
relation we can calculate the values for $n$$_e$ and -- by extension -- $n$$_p$ as well as the 
X-ray emitting mass $M$ of the diffuse emission. Another important
physical parameter to consider is the corresponding pressure $P$ of the SNR. If we define the
total number density of particles $n$$_{total}$ to be the sum of the number densities of electrons,
protons and helium nuclei (that is, $n$$_{total}$ = $n$$_e$ + $n$$_p$ + $n$$_{He}$), we can 
make the approximation $n$$_{total}$ $\approx$ 2 $n$$_e$ for a plasma with cosmic abundances.
Therefore, we may express $P$ as 

\begin{equation}
P/k = 2 n_e T~\mbox{K cm$^{-3}$},
\label{PressureEqn}
\end{equation}
where $k$ is Boltzmann's constant and $T$ is the temperature of the SNR expressed in Kelvin.
Finally, we estimate the age $t$ of G352.7$-$0.1 using the same relation employed by 
\citet{Giacani09} (see Section \ref{IntroductionSection}), that is,
\begin{equation}
t \sim \tau/n_e.
\label{tauequation}
\end{equation}
where $\tau$ is the ionization timescale of the X-ray-emitting plasma (see \citet{Borkowski01}). 
\par
We first calculate values for $n$$_e$ for both the ejecta-dominated component and the 
ISM-dominated component using the normalizations obtained when using the 
PHABS$\times$(VNEI+NEI) model to fit the extracted {\it Chandra} spectrum of the whole SNR
(see Table \ref{ChandraPHABSVNEINEISpectralTable}). If we approximate the volume of
G352.7$-$0.1 to be a spheroid with radii of 2.4 arcmin $\times$ 2.1 arcmin $\times$ 2.3 arcmin
(corresponding to radii of 5.2 pc $\times$ 4.6 pc $\times$ 5.0 pc at the assumed distance of
7.5 kpc to the SNR), the corresponding volume of the SNR is $\sim$1.1$\times$10$^{58}$ cm$^{-3}$.
Using Equation \ref{ThermalNormEqn}, we calculate (for the ejecta-dominated component)
an electron density $n$$_e$ = 0.33 cm$^{-3}$, a proton density $n$$_p$ = 0.28 cm$^{-3}$ and an 
X-ray emitting mass of $M$$_X$ = $n$$_e$ $\times$ $n$$_p$ $\times$ $V$ = 2.6 $M$$_{\odot}$. 
Applying the same equation to the ISM-dominated component, we calculate $n$$_e$ = 6.0 
cm$^{-3}$, $n$$_p$ = 4.9 cm$^{-3}$ and $M$$_X$ = 45 $M$$_{\odot}$. While the calculated X-ray
emitting mass for the ejecta-dominated component of the SNR is comparable to that seen for other
Galactic SNRs, the X-ray emitting-mass for the ISM-dominated component of the SNR is remarkably
large. For comparison purposes, \citet{Rho02} estimated that the swept-up X-ray-emitting mass
associated with the MM SNR W28 was only $\sim$20-25 $M$$_{\odot}$. From 
Equation \ref{PressureEqn}, we calculate corresponding pressures of the 
ejecta-dominated component and the ISM-dominated components to be 2.5$\times$10$^7$
cm$^{-3}$ K and 3.3$\times$10$^7$ cm$^{-3}$ K, respectively. These two values are nearly
three orders of magnitude greater than the typical ISM pressure. We then use Equation
\ref{tauequation} to estimate the age of the
SNR to be $t$ $\sim$ 1.7$\times$10$^{10}$ cm$^{-3}$ s / 0.33 cm$^{-3}$ $\sim$ 1600 years. 
Finally, we estimate the initial explosion energy $E$$_{SN}$ of this supernova using the well-known
Sedov equation \citep{Sedov59}
\begin{equation}
E_{SN} = \frac{r^5 \rho}{t^2}.
\end{equation}
Assuming an age $t$=1600 years, a radius $r$=4.9 pc and
a mass density $\rho$ = $n$$_p$$m$$_p$ (where $n$$_p$ = 4.6 cm$^{-3}$ and $m$$_p$
is the mass of a proton), we calculated $E$$_{SN}$ = 2.54$\times$10$^{51}$ ergs.
\par 
We now compare our computed physical parameters to those determined by \citet{Giacani09}.
Our estimate of the electron number density $n$$_e$ for the ejecta-dominated component is quite 
similar to the value derived for that parameter by those authors. Our estimate of the swept-up
mass is higher than that calculated by \citet{Giacani09} while our age estimate is less than half the
age derived by those authors and is comparable to the age estimate given by \citet{Kinugasa98}. 
Our estimate of the initial explosion energy is also an order of magnitude greater than the estimates
of the initial explosion energy provided by \citet{Giacani09} and \citet{Kinugasa98}.
We attribute differences 
in part to our use of two thermal components while 
fitting the extracted {\it Chandra} spectra while \citet{Giacani09} only used a single
component to fit the extracted {\it XMM-Newton} spectra. 

\subsection{Is the X-ray Emitting Plasma Associated With G352.7$-$0.1 Overionized?}
\label{OverionizationSubSection}

We now investigate whether the X-ray-emitting plasma associated with G352.7$-$0.1 is
overionized. Here, overionization means that atomic species are observed to be in a higher
ionization state than expected based on the observed electron temperature of the X-ray-emitting
plasma. Several Galactic MMSNRs are known to feature overionized X-ray-emitting
plasmas: examples of such MMSNRs include IC 443 \citep{Yamaguchi09}, W49B
\citep{Ozawa09}, G346.6$-$0.2 \citep{Yamauchi12} and G359.0-0.1 \citep{Ohnishi11}. 
Reviews of the phenomenon of overionized X-ray-emitting plasmas associated with
SNRs is provided by \citet{Miceli11} and \citet{Vink12}.
\par
To investigate whether or not the X-ray-emitting plasma associated with G352.7$-$0.1 is
indeed over ionized, we follow the analysis presented by \citet{Yamauchi12}. Using the SPEX 
software package \citep{Kaastra96}, those authors fitted the {\it Suzaku} XIS spectra of the SNR 
G346.6$-$0.2 using the {\tt neij} model. According to this model (see \citet{Yamauchi12}), an X-ray 
emitting plasma may be described such that the initial ion temperature $kT$$_Z$ and the original 
electron temperature $kT$$_{e1}$ were the same (that is, $kT$$_Z$ = $kT$$_{e1}$) and the 
plasma itself was in collisional ionization equilibrium. Over time, the electron temperature drops
to $kT$$_{e2}$ due to rapid electron cooling. Such a model can reflect accurately a rapid 
electron cooling as predicted in models of SNR evolution.
\par
We used the SPEX (Version 2.03.03) tool {\tt trafo} 
(Version 1.02) to convert the extracted {\it Chandra} spectral files for the whole SNR into 
{\tt .spo} files suitable for analysis by the software package. 
We attempted to fit with the {\tt neij} model by adopting a value of 5 keV for the initial electron
temperature $kT$$_{e1}$ (similar to the procedure taken by \citet{Yamauchi12}), freezing 
$kT$$_{e1}$ to this value and thawing the abundances of
silicon and sulfur while allowing the other fit parameters ($N$$_H$, normalization, $kT$$_{e2}$,
$kT$$_Z$ and $\tau$) to vary. We were unable to generate a statistically-acceptable fit using the
{\tt neij} model that is superior (that is, with a lower value for $\chi$$^2_{\nu}$) than that obtained
for the thermal models described above. We therefore conclude that
the X-ray emitting plasma associated with G352.7$-$0.1 is not overionized.

\subsection{Search for Unresolved X-ray Sources}
\label{UnresolvedXraySourceSubSection}

We next conducted a search for a neutron star associated with G352.7$-$0.1, taking
advantage of the high angular resolution capabilities of {\it Chandra}. To the best of our
knowledge, it is the first search for the neutron star in G352.7$-$0.1.
As noted above, G352.7$-$0.1 is suspected to be interacting with molecular clouds based on the
detection of maser emission: such a result suggests (but does not prove) that the stellar 
progenitor of G352.7$-$0.1 was formed near a molecular cloud and is massive. Another robust
argument for a massive stellar progenitor associated with this SNR is the large calculated swept-up
mass of 45 $M$$_{\odot}$ coupled with the X-ray-emitting ejecta mass of 2.6 $M$$_{\odot}$. 
Therefore, a search
for the neutron star produced in the same supernova explosion that parented this SNR is warranted.
There is also a possibility that the supernova explosion that produced G352.7$-$0.1 may also have
fostered a black hole (see e.g., \citet{Kaplan04,Kaplan06} for the discussion of the non-detections of 
compact objects in several other SNRs). 
\par
\begin{figure}
\plotone{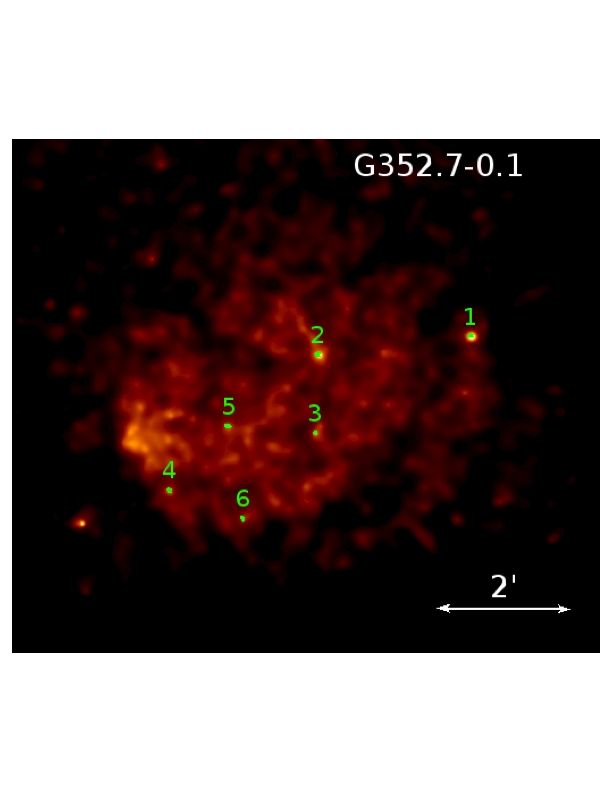}
\caption{A hard-band (2-10 keV) {\it Chandra} image of G352.7$-$0.1 with the locations of the point 
sources detected using {\tt wavdetect} marked and labeled. See Section 
\ref{UnresolvedXraySourceSubSection} and Table \ref{G352unresolvedTable}.\label{G352HardSourcesImage}}
\end{figure}

To conduct our search, we used the CIAO tool $\tt{dmcopy}$ to create an event list and a ``hard" 
image for the energy range from 2.0 to 10.0 keV.  We then ran the CIAO tool 
$\tt{wavdetect}$ -- a wavelet-based unresolved source detection tool \citep{Freeman02} --
with its default parameters on this image to find unresolved sources seen within the apparent angular 
extent of G352.7$-$0.1. In Figure \ref{G352HardSourcesImage} we present our hard image of
G352.7$-$0.1 with the locations of the unresolved sources detected by {\tt wavdetect} indicated. Six 
unresolved sources 
were detected by this tool at a minimum of a S/N$\gtrsim$ 4 level and properties of these sources 
are listed in Table \ref{G352unresolvedTable}. For the two brightest detected sources, we performed 
spectral fitting in XSPEC using an absorbed power law model with $N$$_H$ fixed at 
2.6$\times$10$^{22}$ cm$^{-2}$ (the average Galactic value in this direction for d$\sim$7.5 kpc). 
Due to the small number of counts we did not bin the data and used C-statistics while performing 
the fits. Sources 1 and 2 both show extremely hard (albeit uncertain) 
spectra with $\Gamma=-0.4\pm0.3$ and  $\Gamma=-0.5\pm0.4$, respectively. If $N$$_H$ is 
allowed to be free, its value increases by a factor of two compared to the frozen value (for both 
sources) and the photon 
indices become  $\Gamma\sim0.3$ and  $\Gamma\sim0.1$, respectively. These parameters are, 
however, poorly constrained due to the small total number of counts detected from each source,
especially at energies below 2 keV. Since it is unlikely for the Galactic $N$$_H$ to be even 
larger\footnote{The average Galactic HI column density in this direction is (1.4--1.6)$\times10^{22}$ 
cm$^{-2}$ -- see \citet{Dickey90}.}, the spectra of these sources must also be intrinsically hard. Such 
hard spectra are unusual for young isolated neutron stars or pulsars, while the large X-ray to 
near-infrared (K-band) flux ratio ($\gtrsim$30 in K-band) implies that both sources are likely to be 
either heavily obscured (that is, the soft component is completely absorbed) magnetic cataclysmic 
variables (CVs) in the hard state, obscured remote high mass X-ray binaries (HMXBs) in the 
low-hard state or AGNs. The rest of the sources in Table \ref{G352unresolvedTable} have even fewer 
counts which precludes any spectral analysis in the X-ray. 
Sources 5 and 6 have counterparts detected in the near-infrared (by 2MASS) and/or infrared 
(by the {\sl Spitzer} GLIMPSE survey), therefore neither of them 
can be an isolated pulsar or neutron star related to G352.7$-0.1$. Finally, very little information is 
available about sources 3 and 4. The lack of near-infrared and infrared counterparts for these 
sources coupled with their hard band X-ray  
luminosities ($\simeq 10^{32}$ erg s$^{-1}$ at d = 7.5 kpc)  are in principle consistent with those of  
isolated pulsars at this distance. Indeed, as shown in the top panel in Figure 3 from 
\citet{Kargaltsev13}, the scatter in the X-ray 
efficiencies ($\eta_{X}=L_X/\dot{E}$) of young (characteristic ages $t\lesssim 10$ kyr; spin-down 
energy-loss rates $\dot{E}\gtrsim10^{37}$ erg s$^{-1}$) pulsars is nearly 5 orders of magnitude. 
Therefore,  $L_X\simeq 10^{32}$ ergs s$^{-1}$ can be accommodated by assuming low efficiency 
$\eta_{X}\lesssim10^{-4}$ (although the number of known pulsars exhibiting such low efficiencies is 
fairly small). Alternatively, Sources 3 and 4 may be remote faint AGNs whose soft X-ray and 
optical and near-infrared emission is severely attenuated by the large absorbing column along this 
line of sight. 
\par
We also comment on the possibility that a compact central object (CCO) could be associated with 
G352.7$-$0.1. CCOs are a type of isolated NSs which are found in the centers of their host SNRs. 
They exhibit purely thermal emission (with surface temperatures of $kT$$\sim$ $0.2$--$0.4$ keV) 
and feature luminosities of 
$L$$_X$ $\times10^{33}$ erg s$^{-1}$ over the 0.5 to 8 keV energy range\citep{deLuca08}. 
Therefore, Source 3 could be an absorbed CCO associated with G352.7$-$0.1. Alternatively, if 
G352.7$-0.1$ is further away than the assumed 7.5 kpc, the CCO (if it exists) may not be detectable  
in the existing {\it XMM-Newton} and {\sl Chandra} datasets. 
\par
Since none of these six sources appears to be a particularly promising candidate for a compact 
object associated with G352.7$-0.1$, there remains an open question about the presence of a 
compact object leftover from the SN explosion (assuming it was indeed a core-collapse supernova 
and not a Type Ia supernova). The remaining option would be an isolated black hole which can be
difficult to detect because of the very low accretion rate (assuming Bondi-Hoyle accretion) inside the 
rarefied hot gas filling the interior of the SNR \citep{Beskin05}. 
\par
We also note that none of the sources detected by {\tt wavdetect} at a signal-to-noise ratio of
four or greater are physically coincident
with the radio sources WBH2005 352.775-0.153 and WBH2005 352.772-0.149 seen toward the
eastern rim of the SNR and speculated to be associated with a distant AGN (see Section 
\ref{IntroductionSection}). We therefore place an upper limit of $\sim$10$^{-16}$ ergs cm$^{-2}$ 
s$^{-1}$ on the absorbed X-ray fluxes from these two radio sources.

\subsection{Infrared Emission from G352.7$-$0.1}
\label{IRG352SubSection}

\begin{figure}
\includegraphics[angle=90,scale=0.65]{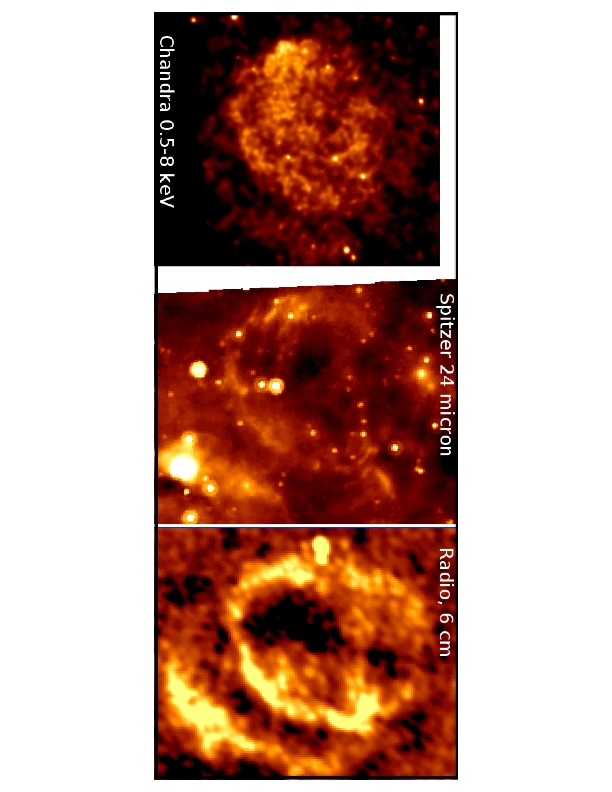}
\label{G352MultiwavelengthFigure}
\caption{Multiwavelength images of G352.7$-$0.1. {\it (right)} {\it Chandra} broadband X-ray image (0.5
- 8.0 keV) smoothed with a Gaussian with a 1.5 arcseconds. {\it (center)} {\it Spitzer} MIPS
24 micron image. {\it (right)} {\it VLA} 6 cm image. Notice the contrasting X-ray and radio 
morphologies of the SNR (typical of MMSNRs) and the robust morphological
similarities between the infrared and radio images. See Section \ref{IRG352SubSection}.}
\end{figure}

We have searched for an infrared counterpart to G352.7$-$0.1 by examining
archival observations of the field that includes this SNR. We find a counterpart from data
collected at a wavelength of 24$\mu$m made with the Multiband Imaging Photometer (MIPS --
\citet{Rieke04}) aboard the {\it Spitzer Space Telescope} \citep{Werner04} as part of the 
MIPSGAL survey that sampled the inner Galactic plane at the wavelengths of 24$\mu$m and
70 $\mu$m \citep{Carey09}. In Figure \ref{G352MultiwavelengthFigure}, we present three images 
of G352.7$-$0.1 at different wavelengths -- namely a broadband (0.5 - 8.0 keV) X-ray image made 
with {\it Chandra}, an infrared 24$\mu$m image made with MIPS aboard {\it Spitzer} and a
radio continuum 6cm map made with the VLA. As we previously mentioned, one can clearly see the
contrasting center-filled X-ray morphology and shell-like radio morphology that categorize 
G352.7$-$0.1 as an MMSNR. Also note in these images the strong correlation between the regions of
infrared emission and regions of radio emission. At both wavelengths, a nearly complete inner
ring of emission is seen along with an outer incomplete ring of emission seen toward the southeast.
A study of fourteen Galactic SNRs detected by MIPS was presented by \citet{Andersen11}. Those 
authors found evidence for interactions between these SNRs and adjacent molecular clouds
based on the detection of [O I] emission ionic lines and emission from molecular lines. We suspect
that the observed flux as detected by MIPS from G352.7$-$0.1 probably has a very similar origin. 
We also note
that -- like all of the sources considered in the survey conducted by \citet{Andersen11} -- G352.7$-$0.1
is an MMSNR. We therefore argue that the detection of infrared emission from this
SNR helps strengthen the link between MMSNRs and SNRs interacting with adjacent 
molecular clouds. In a recent work \citep{Pannuti14}, we found additional evidence for this link in a
study of the X-ray properties of four Galactic MMSNRs that were detected by {\it Spitzer} at shorter
infrared wavelengths (such as 4.5 $\mu$m) with the Infrared Array Camera (IRAC -- \citet{Fazio04})
in the survey conducted by \cite{Reach06}. 

\subsection{G352.7$-$0.1: An Ejecta-Dominated Mixed-Morphology SNR}
\label{EjectaDominatedMMSNRSubSection}

In Table \ref{EjectaDominatedMMSNRsTable} we provide a list of known Galactic 
MMSNRs which are known to feature ejecta-dominated X-ray emission.
By ``ejecta-dominated," we mean that the measured abundances of certain elements (metals) are
elevated relative to solar abundances. Examples of such elements are silicon and sulfur,
which are seen to be elevated in the X-ray-emitting plasmas of MMSNRs: in Table 
\ref{EjectaDominatedMMSNRsTable} we list examples of sources of this type. As shown in this paper
and in previous analyses conducted by \citet{Kinugasa98} and \citet{Giacani09}, G352.7$-$0.1
(similar to HB 21 and G359.1$-$0.1) features silicon and sulfur abundances that exceed solar. 
\par
The origin of ejecta-dominated emission from MMSNRs
remains elusive. Estimates of the ages of MMSNRs often hover near 
$\sim$10$^4$ years. For such ages, the X-ray-emitting plasma associated with an SNR
is expected to be dominated by the swept-up ISM rather than ejecta with elemental abundances
comparable to solar. 
Ejecta-dominated MMSNRs appear to represent a new evolutionary scenario
through which sources of this type are formed. As summarized by \citet{Vink12}, a high metal
content in the X-ray emitting plasma may help explain the observed emission characteristics of
MMSNRs for two reasons. Firstly, metal-rich plasmas produce more X-ray emission than plasmas 
which are not metal-enhanced: this leads to a contrast between the X-ray emitting metal-rich interior 
and the metal-poor swept-up exterior of the SNR. Secondly,  
the uniform temperatures in the interiors of MM SNRs can be
explained using thermal conduction models \citep{Cox99}: this explanation is challenged by 
magnetic fields limiting thermal conduction along magnetic field lines, but if the interiors of MM SNRs
are dominated by ejecta, then the stellar magnetic field may be reduced due to the expansion and is 
thus significantly weakened (to values as low as 10$^{-14}$ G -- see \citet{Vink12}). At such a
low value conduction across field lines becomes as efficient as conduction along field lines: thus
thermal conduction becomes efficient in these conditions but only to the boundaries between the
ejecta-dominated matter and swept-up matter. At these boundaries, Rayleigh-Taylor instabilities
may contain magnetic fields with elevated strengths, thus hampering efficient thermal conduction.
\par 
Of the SNRs listed in Table \ref{EjectaDominatedMMSNRsTable}, the only two which may possibly
host neutron stars are CTB 1 \citep{Pannuti10} and IC 443. However, in the case of IC 443,
the observed neutron star is unlikely to be associated with the SNR because it is observed to move
in the wrong direction (that is, toward the center of the SNR). Therefore, it may be that MMSNRs
are such that after the explosion, their progenitors do not leave a neutron star: either a black hole
is produced or nothing is left. If the ejecta mass is large, then it is likely that the the progenitor
star was a massive star and a black hole was left behind. Such progenitors may explain the
elevated abundances in MMSNRs and the fact that they are close to molecular clouds (where
massive stars are formed more easily). 
\par
We have searched for other characteristics that the SNRs listed in Table 
\ref{EjectaDominatedMMSNRsTable} may have in common. For example, we considered whether
the radio morphologies of these SNRs are all similar, such as bilateral (also known as 
``barrel-shaped" -- see \citet{Kesteven87}): examples of well-known barrel-shaped Galactic SNRs 
include
G296.5$+$10.0 \citep{Storey92} and G320.4$-$1.2 \citep{Dubner02}. In fact, in a seminal study 
of Galactic SNRs presented by \citet{Kesteven87} that classified these sources based on their radio 
morphologies did indeed identify one of the SNRs in Table \ref{EjectaDominatedMMSNRsTable} --
namely CTB 1 -- as a ``well-developed" barrel: other SNRs like W44, HB21, HB3, IC 443, and
Kes 27 were described as ``difficult to classify" by those authors. 
More recently, \citet{Keohane07} argued that the SNR W49B should be classified 
as a barrel-shaped SNR (it had been classified as a ``possible barrel" by \citet{Kesteven87}). 
More detailed radio observations and analysis need to be conducted of these SNRs (and the
sample of known ejecta-dominated mixed-morphology SNRs must be increased) to determine
if barrel-shape morphology is consistently seen amongst these sources.

\section{Conclusions}
\label{ConclusionsSection}

The conclusions of this paper may be summarized as follows:
\par
1) Complementary {\it XMM-Newton} and {\it Chandra} observations of the Galactic SNR
G352.7$-$0.1 confirm a center-filled thermal morphology with a contrasting shell-like radio
morphology. These observations motivate a classification of this source as a MMSNR.\\
2) Analysis of extracted {\it XMM-Newton} spectra for the whole SNR and for the bright eastern
region can be fit satisfactorily with a single thermal component that describes an X-ray-emitting 
plasma that is not in collisional ionization equilibrium with enhanced abundances of silicon and sulfur. 
This result is consistent with previous analysis of extracted {\it ASCA} and {\it XMM-Newton} spectra of 
this SNR. In contrast, analysis (presented here for the first time) of extracted {\it Chandra} spectra of
the whole SNR as well as separate regions of the SNR cannot all be fit satisfactorily with a single
thermal component. In the cases of the ``left" and ``top" regions as well as the whole SNR, 
statistically-acceptable fits are obtained when either a second thermal component with solar
abundances is added or when two thermal components with different temperatures (one with 
enhanced silicon abundance and the other with enhanced sulfur abundance). We argue that the
former scenario is more physically plausible.\\
3) We have calculated physical parameters (computed based on our derived fit parameters) for
G352.7$-$0.1, include $n$$_e$, $n$$_p$, $P$, $M$ and $t$. Most notably, our computed masses
of the X-ray-emitting masses of the ejecta-dominated component and the ISM-dominated component
are 2.6 $M$$_{\odot}$ and 45 $M$$_{\odot}$, respectively. This is a remarkably high swept-up mass 
for a Galactic SNR and may indicate a unique evolutionary scenario (involving a massive progenitor
star interacting with a dense molecular cloud environment) for this SNR. \\
4) We have conducted a spectral analysis of the extracted {\it Chandra} spectrum of the whole SNR
to determine if the X-ray emitting plasma associated with this SNR is over-ionized. We cannot obtain
a satisfactory fit to the extracted spectrum using the {\tt neij} model with parameters that might indicate
over-ionization: we therefore conclude that the X-ray emitting plasma associated with G352.7$-$0.1
is not over-ionized. \\
5) We have searched for a hard unresolved X-ray source that may be the neutron star associated with
G352.7$-$0.1. We find six unresolved X-ray sources seen in projection toward the interior of the SNR but 
none of these sources may be firmly classified as a neutron star physically 
associated with G352.7$-$0.1. We also find no X-ray counterparts to the background radio galaxy 
(with prominent radio lobes) seen in projection toward the eastern rim of the SNR. \\
6) We have presented for the first time the detection of infrared emission (namely emission detected
at 24 $\mu$m by MIPS aboard {\it Spitzer}) from this SNR. The infrared morphology is shell-like and 
strongly resembles the radio morphology, including the complete inner radio shell and incomplete
outer radio shell. Such emission has been detected previously from SNRs known to be interacting with 
adjacent molecular clouds: this detection helps confirm that G352.7$-$0.1 is indeed interacting with an 
adjacent molecular cloud. \\
7) We have compared G352.7$-$0.1 to other ejecta-dominated MMSNRs. The origin of 
ejecta-dominated emission from MMSNRs remains unknown and may indicate a new 
scenario for the evolution of this class of source. 

\acknowledgments

We thank the referee for helpful comments and suggestions that have improved the quality of
this manuscript. This research has been supported by a grant from the Kentucky Space Grant
Consortium. O.K. acknowledges support from archival Chandra/NASA award AR3-14017X.
T.G.P. thanks Jelle de Plaa for his assistance with installing and using the SPEX
software package: he also thanks Keith Arnaud for helpful assistance in the use of the XSPEC
software package. 
T.G.P. also thanks David Alex May and Tiffany Murray for their assistance in
reducing the {\it Chandra} data for G352.7$-$0.1. We also thank Gloria Dubner for kindly sharing
her radio maps of G352.7$-$0.1 with us. This research has made use of 
NASA's Astrophysics Data System and the NASA/IPAC Extragalactic Database (NED)
which is operated by the Jet Propulsion Laboratory, California Institute of Technology,
under contract with the National Aeronautics and Space Administration. This research
has also made use of data obtained from the {\it Chandra Data Archive} and software
provided by the {\it Chandra X-ray Center} (CXC) in the application packages {\it CIAO} and 
{\it ChIPS.} Finally, this work is based on observations obtained with {\it XMM-Newton}, an
ESA science mission with instruments and contributions directly funded by ESA Member States
and the USA (NASA).

\clearpage

\begin{deluxetable}{lcc}
\tablecaption{General Properties of G352.7$-$0.1\label{G352PropsTable}}
\tabletypesize{\scriptsize}
 \tablewidth{0pt}
 \tablehead{\colhead{Property} &  \colhead{Value} & \colhead{References} }
 \startdata
R.A. (J2000.0) & 17 27 40 & (1) \\
Decl. (J2000.0)	& $-$35 07 & (1) \\ 
Angular Diameters (arcmin$\times$arcmin) & 8$\times$6 & (1) \\
Distance (kpc)  & 7.5$\pm$0.5 & (2) \\
Physical Diameters (pc) & 17$\times$13 & (3) \\
Flux Density at 5000 MHz (Jy) & 2.3 & (4) \\ 
Flux Density at 1465 MHz (Jy) & 3.4$\pm$0.4 & (5) \\
Flux Density at 1415 MHz (Jy) & 4.7 & (6) \\
Flux Density at 1400 MHz (Jy) & 3.1$\pm$0.3 & (7) \\
Flux Density at 843 MHz (Jy) & 4.4 & (8) \\
Flux Density at 408 MHz (Jy) & 9.6 & (4) \\
Spectral Index $\alpha$ ($S$$_{\nu}$ $\propto$ $\nu$$^{\alpha}$) & $-$0.6 & (5)   
\enddata
\tablecomments{References: (1) -- \citet{Green09a}, (2) -- \citet{Giacani09}, (3) -- This paper,
(4) -- \citet{Clark75}, (5) -- \citet{Dubner93}, (6) -- \citet{Caswell83}, (7) -- \citet{Giacani09},
(8) -- \citet{Whiteoak96}.}
\end{deluxetable}

\clearpage

\begin{deluxetable}{cccccccccc}
\tablewidth{0pt}
\tabletypesize{\scriptsize}
\rotate
\tablecaption{Summary of {\it XMM-Newton} MOS1, MOS2 and PN Observations of 
G352.7$-$0.1\label{XMMObsTable}}
\tablewidth{0pt}
\tablehead{
& & & & \colhead{MOS1} & & \colhead{MOS2} & & & \colhead{PN}\\
& & & & \colhead{Effective} & \colhead{MOS1} & \colhead{Effective} & \colhead{MOS2} &
\colhead{PN} & \colhead{Effective}\\
& & && \colhead{Exposure} & \colhead{Count} & \colhead{Exposure} 
& \colhead{Count} & \colhead{Exposure} & \colhead{Count} \\
\colhead{Sequence} & \colhead{Observation} & \colhead{R.A.}  & \colhead{Decl.}
& \colhead{Time} &  \colhead{Rate} & \colhead{Time} & \colhead{Rate}
& \colhead{Time}  & \colhead{Rate} \\
\colhead{Number} & \colhead{Date} & \colhead{(J2000.0)} 
& \colhead{(J2000.0)} & \colhead{(s)} & \colhead{(counts s$^{-1}$)} & \colhead{(s)} & 
\colhead{(counts s$^{-1}$)} & \colhead{(s)} & \colhead{(counts s$^{-1}$)}
}
\startdata
0150220101 & 2002 October 3 & 17 27 35.0 & -35 07 23 & 28639 & 9.2$\times$10$^{-2}$ & 28650 
& 9.3$\times$10$^{-2}$ & 16468 & 2.3$\times$10$^{-1}$
\enddata
\tablecomments{The units of Right Ascencion are hours, minutes and seconds while the
units of Declination are degrees, arcminutes and arcseconds. Count rates are for the entire angular 
extent of the diffuse emission and correspond to the energy range from 1.0 keV to 5.0 keV.} 
\end{deluxetable}

\begin{deluxetable}{cccccccc}
\tabletypesize{\scriptsize}
%\rotate
\tablecaption{Summary of {\it Chandra} ACIS-S Observation of 
G352.7$-$0.1\label{ChandraObsTable}}
\tablewidth{0pt}
\tablehead{
& & & & & & \colhead{ACIS-S} & \colhead{ACIS-S} \\
\colhead{Sequence} & & \colhead{Observation} & \colhead{R.A.} & \colhead{Decl.} & 
\colhead{Roll} & \colhead{Effective Exposure} & \colhead{Count} \\
\colhead{Number} & \colhead{ObsID} & \colhead{Date} & \colhead{(J2000.0)} & 
\colhead{(J2000.0)} & \colhead{(deg)} & \colhead{Time (s)} & \colhead{Rate (s)}
}
\startdata
500471 & 4652 & 2004 October 6 & 17 27 41.0 & $-$35 06 45 & 263 & 44583 & 
1.4$\times$10$^{-1}$
\enddata
\tablecomments{The units of Right Ascencion are hours, minutes and seconds while the
units of Declination are degrees, arcminutes and arcseconds. Count rates are for the entire
angular extent of the diffuse emission and correspond to the energy range from 1.0 keV to 5.0 keV.}
\end{deluxetable}

\clearpage

\begin{deluxetable}{lcc}
\tabletypesize{\scriptsize}
\tablecaption{Summary of {\it XMM-Newton} MOS1+MOS2+PN Spectral Analysis of 
G352.7$-$0.1 (PHABS$\times$VNEI Model)\label{XMMSpectralTable}}
\tablewidth{0pt}
\tablehead{
& \colhead{Eastern} & \colhead{Whole} \\
\colhead{Parameter} & \colhead{Region} & \colhead{SNR} 
}
\startdata
$N$$_H$ (10$^{22}$ cm$^{-2}$) & 2.46$^{+0.56}_{-0.46}$ & 2.78$^{+0.24}_{-0.28}$\\
$kT$ (keV) & 1.90$^{+1.60}_{-0.70}$ & 1.20$^{+0.34}_{-0.24}$ \\
$\tau$ (10$^{10}$ cm$^{-3}$ s) & 3.57$^{+3.43}_{-1.57}$& 4.07$^{+2.53}_{-1.17}$  \\
Si & 2.6$^{+1.5}_{-0.9}$ & 2.3$^{+0.6}_{-0.4}$ \\
S & 4.5$^{+3.5}_{-1.7}$ & 3.5$^{+1.4}_{-0.8}$ \\
Normalization (cm$^{-5}$) & 4.9$\times$10$^{-4}$ & 3.5$\times$10$^{-3}$ \\
$\chi$$_{\nu}^2$ ($\chi$$^2$/DOF) & 1.19 (84.58/71) & 1.15 (444.17/387) \\
Absorbed Flux (ergs cm$^{-2}$ s$^{-1}$) &  2.4$\times$10$^{-13}$ & 8.4$\times$10$^{-13}$ \\
Unabsorbed Flux (ergs cm$^{-2}$ s$^{-1}$) & 1.3$\times$10$^{-12}$ & 6.4$\times$10$^{-12}$ \\
Unabsorbed Luminosity (ergs s$^{-1}$) & 8.8$\times$10$^{33}$ & 4.3$\times$10$^{34}$ 
\enddata
\tablecomments{All stated error bounds correspond to 90\% confidence intervals. Elemental 
abundances
are expressed with respect to solar abundances. Normalizations are defined as 
(10$^{-14}$/4$\pi$$d$$^2$ 
$\int$ $n$$_e$ $n$$_H$ dV, where $d$ is the distance to G352.7$-$0.1 in cm, $n$$_e$ and 
$n$$_p$ are the electron and hydrogen number densities, respectively (both in cm$^{-3}$) and
$V$= $\int$ dV is the total volume of the emitting region (in cm$^3$).  Fluxes and luminosities 
are expressed 
for the energy range from 1.0 to 5.0 keV. Luminosities are calculated assuming a distance of
7.5 kpc to G352.7$-$0.1. See Section \ref{XMMG352}.}
\end{deluxetable}

\clearpage

\begin{deluxetable}{lccccccc}
%\rotate
\setlength{\tabcolsep}{0.02in}
\tabletypesize{\scriptsize}
\tablecaption{Summary of {\it Chandra} ACIS Spectral Analysis of 
G352.7$-$0.1 (PHABS$\times$VNEI Model) \label{ChandraPHABSVNEISpectralTable}}
\tablewidth{0pt}
\tablehead{
& \colhead{Bottom} & \colhead{Right} & \colhead{Middle} & \colhead{Crescent} & 
\colhead{Left} & \colhead{Top} & \colhead{Whole} \\
\colhead{Parameter} & \colhead{Region} & \colhead{Region} & \colhead{Region}  
& \colhead{Region}  
& \colhead{Region} & \colhead{Region} & \colhead{SNR}
}
\startdata
$N$$_H$ (10$^{22}$ cm$^{-2}$) &  3.08$^{+0.70}_{-0.60}$ & 
3.10$^{+0.95}_{-0.74}$ & 2.68$^{+0.76}_{-0.64}$ & 2.31$^{+1.93}_{-0.91}$ & 
2.72$^{+0.50}_{-0.44}$ & 3.24$^{+0.72}_{-0.78}$ & 2.71$^{+0.39}_{-0.33}$ \\
$kT$ (keV) &  0.80$^{+0.75}_{-0.25}$ & 1.23$^{+1.93}_{-0.52}$ & 
1.24$^{+1.30}_{-0.56}$ & 2.03 ($>$0.54) & 1.38$^{+0.82}_{-0.50}$ & 0.68$^{+0.82}_{-0.24}$ & 
1.62$^{+0.80}_{-0.46}$  \\
$\tau$ (10$^{10}$ cm$^{-3}$ s) & 12.7 ($>$2.00) & 2.74$^{+12.0}_{-1.21}$ 
& 1.68$^{+1.42}_{-0.62}$ & 1.04$^{+1.10}_{-0.54}$ & 2.46$^{+1.34}_{-0.66}$ & 6.11 ($>$2.44) & 
2.01$^{+0.85}_{-0.41}$ \\
Si & 2.4$^{+1.1}_{-0.7}$  & 2.3$^{+2.9}_{-1.2}$ & 3.4$^{+2.8}_{-1.5}$ & 4.8 ($>$1.2) &
2.2$^{+1.3}_{-0.8}$ & 2.5$^{+2.7}_{-0.9}$ & 1.9$^{+1.3}_{-0.8}$ \\
S & 5.5$^{+2.3}_{-1.8}$  & 3.8$^{+3.6}_{-1.3}$ & 8.1$^{+2.3}_{-3.7}$ & 9.8 ($>$2.0) & 
6.6$^{+3.4}_{-2.9}$ & 6.5$^{+6.5}_{-3.5}$ & 3.6$^{+4.2}_{-1.6}$ \\
Normalization (cm$^{-5}$)  & 1.2$\times$10$^{-3}$ & 6.2$\times$10$^{-4}$ & 
7.3$\times$10$^{-4}$ & 1.8$\times$10$^{-4}$ & 8.7$\times$10$^{-4}$ & 1.6$\times$10$^{-3}$ 
& 3.6$\times$10$^{-3}$ \\
$\chi$$_{\nu}^2$ & 1.07 & 1.06 & 1.04 & 1.02 & 1.24 & 1.25 & 1.28 \\
$\chi$$^2$/DOF & 70.80/66 & 63.83/60 & 78.22/75 & 70.41/69 & 104.20/84 & 76.38/61 
& 266.55/209 \\
Absorbed Flux (ergs cm$^{-2}$ s$^{-1}$)  & 1.3$\times$10$^{-13}$ & 
1.2$\times$10$^{-13}$ & 1.7$\times$10$^{-13}$ & 8.6$\times$10$^{-14}$ & 2.4$\times$10$^{-13}$ 
& 1.1$\times$10$^{-13}$ & 1.1$\times$10$^{-12}$ \\
Unabsorbed Flux (ergs cm$^{-2}$ s$^{-1}$)  & 1.4$\times$10$^{-12}$ & 
1.1$\times$10$^{-12}$ & 1.3$\times$10$^{-12}$ & 4.3$\times$10$^{-13}$ & 
1.7$\times$10$^{-12}$ & 
1.5$\times$10$^{-12}$ & 7.6$\times$10$^{-12}$ \\
Unabsorbed Luminosity (ergs s$^{-1}$)  & 9.4$\times$10$^{33}$ & 
7.4$\times$10$^{33}$ & 8.8$\times$10$^{33}$ & 2.9$\times$10$^{33}$ & 1.1$\times$10$^{34}$ 
& 1.0$\times$10$^{34}$ & 5.1$\times$10$^{34}$ 
\enddata
\tablecomments{All stated error bounds correspond to 90\% confidence intervals. 
Elemental abundances are expressed with respect to solar abundances. Normalizations are 
defined as (10$^{-14}$/4$\pi$$d$$^2$) 
$\int$ $n$$_e$ $n$$_H$ dV, where $d$ is the distance to G352.7$-$0.1 in cm, $n$$_e$ and 
$n$$_p$ are the electron and hydrogen number densities, respectively (both in cm$^{-3}$) and
$V$= $\int$ dV is the total volume of the emitting region (in cm$^3$).  Fluxes and luminosities 
are expressed 
for the energy range from 1.0 to 5.0 keV. Luminosities are calculated assuming a distance of
7.5 kpc to G352.7$-$0.1. See Section \ref{ChandraG352}.}
\end{deluxetable}

\clearpage

\begin{deluxetable}{lccc}
%\rotate
\tabletypesize{\scriptsize}
\tablecaption{Summary of {\it Chandra} ACIS Spectral Analysis of 
G352.7$-$0.1 (PHABS$\times$(VNEI+NEI) Model) 
\label{ChandraPHABSVNEINEISpectralTable}}
\tablewidth{0pt}
\tablehead{
& \colhead{Left} & \colhead{Top} & \colhead{Whole} \\
\colhead{Parameter} & \colhead{Region} & \colhead{Region} & \colhead{SNR} 
}
\startdata
$N$$_H$ (10$^{22}$ cm$^{-2}$) & 3.75$^{+0.49}_{-0.55}$ & 4.61$^{+0.79}_{-0.77}$ & 
3.18$^{+0.13}_{-0.11}$ \\
$kT$$_{\rm{VNEI}}$ (keV) & 1.18$^{+0.26}_{-0.20}$ & 1.82$^{+2.94}_{-0.66}$ & 
3.24$^{+1.40}_{-0.86}$ \\
Si & 2.1$^{+0.8}_{-0.6}$ & 7.0$\pm$3.0 & 3.4$^{+1.2}_{-0.9}$\\
S & 6.3$^{+2.5}_{-1.9}$ & 17.7$\pm$4.9 & 9.4$^{+3.4}_{-3.0}$ \\
$\tau$$_{\rm{VNEI}}$ (10$^{10}$ cm$^{-3}$ s) & 2.8$^{+1.0}_{-0.6}$ & 2.5$^{+47.5}_{-0.9}$ & 
1.7$\pm$0.3  \\
Normalization$_{\rm{VNEI}}$ (cm$^{-5}$) & 1.42$\times$10$^{-3}$ & 1.93$\times$10$^{-4}$ &
1.58$\times$10$^{-3}$ \\
$kT$$_{\rm{NEI}}$ (keV) & 0.13$^{+0.04}_{-0.02}$ & 0.17$^{+0.11}_{-0.07}$ & 
0.24$^{+0.08}_{-0.05}$ \\
$\tau$$_{\rm{NEI}}$ (10$^{10}$ cm$^{-3}$ s) & 7.4$^{+23.6}_{-5.0}$ & 6.8$^{+73.2}_{-5.2}$ & 
1.4$^{+1.1}_{-0.5}$ \\
Normalization$_{\rm{NEI}}$ (cm$^{-5}$) & 6.6 & 1.7 & 0.5  \\
$\chi$$_{\nu}^2$ ($\chi$$^2$/DOF) & 1.13 (91.65/81) & 1.00 (59.01/59) & 1.07 (219.61/206)\\
Absorbed Flux (ergs cm$^{-2}$ s$^{-1}$) & 2.4$\times$10$^{-13}$ & 1.2$\times$10$^{-13}$ &
1.1$\times$10$^{-12}$ \\
Unabsorbed Flux (ergs cm$^{-2}$ s$^{-1}$) & 7.3$\times$10$^{-12}$ & 8.7$\times$10$^{-12}$ &
1.4$\times$10$^{-11}$ \\
Unabsorbed Luminosity (ergs s$^{-1}$) & 4.9$\times$10$^{34}$ & 5.9$\times$10$^{34}$ & 
9.4$\times$10$^{34}$
\enddata
\tablecomments{All stated error bounds correspond to 90\% confidence intervals. 
Elemental abundances are expressed with respect to solar abundances. Normalizations are 
defined as (10$^{-14}$/4$\pi$$d$$^2$ 
$\int$ $n$$_e$ $n$$_H$ dV, where $d$ is the distance to G352.7$-$0.1 in cm, $n$$_e$ and 
$n$$_p$ are the electron and hydrogen number densities, respectively (both in cm$^{-3}$) and
$V$= $\int$ dV is the total volume of the emitting region (in cm$^3$).  Fluxes and luminosities are 
expressed for the energy range from 1.0 to 5.0 keV. Luminosities are calculated assuming a distance 
of 7.5 kpc to G352.7$-$0.1. See Section \ref{ChandraG352}.}
\end{deluxetable}

\begin{deluxetable}{lccc}
%\rotate
\tabletypesize{\scriptsize}
\tablecaption{Summary of {\it Chandra} ACIS Spectral Analysis of 
G352.7$-$0.1 (PHABS$\times$(VNEI+VNEI) Model) 
\label{ChandraPHABSVNEIVNEISpectralTable}}
\tablewidth{0pt}
\tablehead{
& \colhead{Left} & \colhead{Top} & \colhead{Whole} \\
\colhead{Parameter} & \colhead{Region} & \colhead{Region} & \colhead{SNR} 
}
\startdata
$N$$_H$ (10$^{22}$ cm$^{-2}$) & 2.59$^{+0.14}_{-0.12}$ & 4.22$^{+0.68}_{-0.60}$ & 
2.59$^{+0.27}_{-0.39}$  \\
$kT$$_{\rm{Si}}$ (keV)  & 1.78$^{+0.32}_{-0.30}$ & 0.22$^{+0.08}_{-0.06}$ & 
2.14$^{+3.46}_{-0.84}$ \\
Si & 6.4$\pm$1.5 & 2.9$^{+11.1}_{-1.7}$ & 7 ($>$4) \\
$\tau$$_{\rm{Si}}$ (10$^{10}$ cm$^{-3}$ s) & 0.71$^{+0.22}_{-0.19}$ & 5.8$^{+64.2}_{-4.8}$ & 
0.54$^{+0.16}_{-0.13}$ \\
Normalization$_{\rm{Si}}$ (cm$^{-5}$) & 5.08$\times$10$^{-4}$ & 0.20 & 2.57$\times$10$^{-3}$ \\
$kT$$_{\rm{S}}$ (keV) & 0.61$^{+0.07}_{-0.09}$ & 0.76$^{+0.32}_{-0.28}$ & 0.39$^{+0.33}_{-0.02}$ \\
S & 13.6$^{+11.4}_{-4.1}$ & 9.8($>$4.0) & 37 ($>$7) \\
$\tau$$_{\rm{S}}$ (10$^{10}$ cm$^{-3}$ s) & 103($>$6.8) & 950($>$1.6) & 38 ($>$2.0)\\
Normalization$_{\rm{S}}$ (cm$^{-5}$) & 1.03$\times$10$^{-3}$ & 6.92$\times$10$^{-4}$ &
9.7$\times$10$^{-3}$\\
$\chi$$_{\nu}^2$ ($\chi$$^2$/DOF) & 1.17 (94.48/81) & 1.04 (61.26/59) & 1.05 (215.69/206) \\
Absorbed Flux (ergs cm$^{-2}$ s$^{-1}$) & 2.4$\times$10$^{-13}$ & 1.1$\times$10$^{-13}$ &
1.1$\times$10$^{-12}$ \\
Unabsorbed Flux (ergs cm$^{-2}$ s$^{-1}$) & 1.6$\times$10$^{-12}$ & 5.0$\times$10$^{-12}$ &
7.4$\times$10$^{-12}$ \\
Unabsorbed Luminosity (ergs s$^{-1}$) & 1.1$\times$10$^{34}$ & 3.4$\times$10$^{34}$ & 
5.0$\times$10$^{34}$
\enddata
\tablecomments{All stated error bounds correspond to 90\% confidence intervals. Elemental 
abundances are expressed with respect to solar abundances. Normalizations are defined as 
(10$^{-14}$/4$\pi$$d$$^2$ $\int$ $n$$_e$ $n$$_H$ dV, where $d$ is the distance to G352.7$-$0.1
in cm, $n$$_e$ and $n$$_p$ are the electron and hydrogen number densities, respectively (both in 
cm$^{-3}$) and $V$= $\int$ dV is the total volume of the emitting region (in cm$^3$).  Fluxes and 
luminosities are expressed for the energy range from 1.0 to 5.0 keV. Luminosities are calculated 
assuming a distance of 7.5 kpc to G352.7$-$0.1. See Section \ref{ChandraG352}.}
\end{deluxetable}

\clearpage

\begin{deluxetable}{lccccccccc}

\tablecaption{{unresolved Hard X-ray Sources Detected by {\it Chandra} Toward 
G352.7$-$0.1}\label{G352unresolvedTable}}
\tabletypesize{\scriptsize}
\renewcommand\tabcolsep{3pt}
\tablewidth{0pt}
\tablehead{
& & \colhead{R.A.} & \colhead{Decl.} \\
\# & \colhead{Source}& \colhead{(J2000.0)} & \colhead{(J2000.0)} & \colhead{$N$$_{0.2-10}$} & 
\colhead{$f^{\rm abs}_{0.2-2}$}  &  \colhead{$f^{\rm abs}_{2-10}$} &  HR  &
\colhead{$L_{\rm  abs,2-10}$}  & Type
}
\startdata
1  &  CXOU J172726.9$-$350600 & 17 27 26.97 & $-$35 06 00.2 & 80 &  
 0.1 & 67.5 & 1.00 & 45  & mCV?/qHMXB?/AGN?  \\
2  &  CXOU J172737.7$-$350616 & 17 27 37.74 & $-$35 06 16.3 & 43 &
0   & 64.0  & 1.00 & 43  & mCV?/qHMXB?/AGN?  \\
3  &  CXOU J172737.9$-$350724 & 17 27 37.92 & $-$35 07 24.2 & 28 &
0.9 & 13.3  &  0.87 & 8.9 & mCV?/AGN?/PSR?/CCO?  \\
4  &  CXOU J172748.2$-$350814 & 17 27 48.22 & $-$35 08 14.4 & 16 &
0.1& 19.1 & 0.99 & 13 & mCV?/PSR? \\
5$\tablenotemark{a}$  &  CXOU J172744.0$-$350719 & 17 27 44.05 & $-$35 07 18.9 & 14 &
0& 25.6  &  1.00 & 17 & K/M-dwarf/AGN?  \\
6$\tablenotemark{b}$  &  CXOU J172743.0$-$350839 & 17 27 43.03 & $-$35 08 38.8 & 13 &
1.1 & 14.0  & 0.85 & 9.4 & AGN/PMS-star? 
\enddata
\tablecomments{The units of Right Ascencion are hours, minutes and seconds while the units of 
Declination are degrees, arcminutes and arcseconds. $N$$_{0.2-10}$ is the number of source 
counts  in 0.2--10 keV.  The fluxes ($f^{\rm abs}_{0.2-2}$ and $f^{\rm abs}_{2-10}$ in 0.2--2 and 2--10 
keV, respectively) are absorbed (calculated using the {\em eff2evt} CIAO tool; see 
http://cxc.harvard.edu/ciao/threads/eff2evt/)  in units of $10^{-15}$ ergs cm$^{-2}$ s$^{-1}$. The
hardness ratios ``HR'' are calculated based on the fluxes $f^{\rm abs}_{0.2-2}$ and 
$f^{\rm abs}_{2-10}$, specifically HR = ($f^{\rm abs}_{2-10}$ - $f^{\rm abs}_{0.2-2}$) /
($f^{\rm abs}_{2-10}$ + $f^{\rm abs}_{0.2-2}$).  The luminosities $L$$_{\rm abs,2-10}$ are 
expressed in units of 
$10^{31}$ ergs s$^{-1}$ and are calculated from $f^{\rm abs}_{2-10}$ based on an assumed 
distance of 7.5 kpc to G352.7$-$0.1. The abbreviations for type classification are defined as
follows: mCV = magnetic cataclysmic variable in low-hard state, qHMXB = quasi-high mass X-ray
binary, AGN = active galactic nuclei, PSR = pulsar, CCO = central compact object, PMS-star = 
pre-main sequence star. See Section \ref{UnresolvedXraySourceSubSection}.
}
\tablenotetext{a}{This source has a near-infrared counterpart detected with an apparent K-band
magnitude of 12.0 and an infrared counterpart detected with an apparent 4.5$\mu$m magnitude
of 11.0.}
\tablenotetext{b}{This source has an infrared counterpart detected with an apparent 4.5$\mu$m
magnitude of 12.7.}
\end{deluxetable}

\clearpage

\begin{deluxetable}{lc}
\tablecaption{List of Known Ejecta-Dominated Mixed-Morphology 
SNRs\label{EjectaDominatedMMSNRsTable}}
\tabletypesize{\scriptsize}
\tablewidth{0pt}
 \tablehead{
 \colhead{SNR} & \colhead{Reference}}
 \startdata
W44 (G34.7$-$0.4) & \citet{Shelton04} \\
W49B (G43.3$-$0.2) & \citet{Ozawa09} \\
HB 21 (G89.0$+$4.7) & \citet{Lazendic06}, \citet{Pannuti10}\\
CTB 1 (G116.9$+$0.2) & \citet{Lazendic06}, \citet{Pannuti10} \\
HB 3 (G132.7$+$1.3) & \citet{Lazendic06} \\
IC 443 (G189.1$+$3.0) & \citet{Troja08}, \citet{Yamaguchi09} \\
Kes 27 (G327.4$+$0.4) & \citet{Chen08} \\
G344.7$-$0.1 & \citet{Yamaguchi12} \\
G352.7$-$0.1 & \citet{Kinugasa98}, \citet{Giacani09}, This paper \\
G359.1$-$0.5 & \citet{Ohnishi11}
\enddata
\tablecomments{See Section \ref{EjectaDominatedMMSNRSubSection}.}
\end{deluxetable}

\clearpage

\clearpage

\clearpage

\clearpage

\clearpage

\clearpage

\clearpage

\clearpage

\clearpage

\clearpage

\end{document}